\documentclass[twocolumn,twocolappendix]{aastex631}
\usepackage{newtxtext,newtxmath}

\begin{document}

\title{The Origins of Narrow Spectra of Fast Radio Bursts}

\author{Pawan Kumar}\thanks{E-mail: pk@astro.as.utexas.edu}
\affiliation{Department of Astronomy, University of Texas at Austin, Austin, TX 78712, USA}

\author{Yuanhong Qu}\thanks{E-mail: yuanhong.qu@unlv.edu}
\affiliation{Nevada Center for Astrophysics, University of Nevada, Las Vegas, NV 89154}
\affiliation{Department of Physics and Astronomy, University of Nevada Las Vegas, Las Vegas, NV 89154, USA}

\author{Bing Zhang}\thanks{E-mail: bing.zhang@unlv.edu}
\affiliation{Nevada Center for Astrophysics, University of Nevada, Las Vegas, NV 89154}
\affiliation{Department of Physics and Astronomy, University of Nevada Las Vegas, Las Vegas, NV 89154, USA}

\begin{abstract}
Observations find that some fast radio bursts (FRBs) have extremely narrow-band spectra, i.e., $\Delta\nu/\nu_0 \ll 1$. We show that when the angular size of the emission region is larger than the Doppler beaming angle, the observed spectral width ($\Delta\nu/\nu_0$) exceeds 0.58 due to the {\it high latitude effects} for a source outside the magnetosphere, even when the spectrum in the source's comoving frame is monochromatic. The angular size of the source for magnetospheric models of FRBs can be smaller than the Doppler beaming angle, in which case this geometric effect does not influence the observed bandwidth.
We discuss various propagation effects to determine if any could transform a broad-spectrum radio pulse into a narrow-spectrum signal at the observer's location. We find that plasma lensing and scintillation can result in a narrow bandwidth in the observed spectrum. However, the likelihood of these phenomena being responsible for the narrow observed spectra with $\Delta\nu/\nu_0 < 0.58$ in the fairly large observed sample of FRBs is exceedingly small.
\end{abstract}


\section{Introduction}
Fast Radio Bursts (FRBs) are highly dispersed, bright radio bursts with extremely high brightness temperatures $\sim 10^{36}$ K \citep{Lorimer2007,Petroff2016} which implies the intrinsic radiation mechanisms must be coherent. 
Coherent radiation mechanisms can be generally attributed to two classes within the FRB context: maser and the antenna mechanism. However, the intrinsic radiation mechanisms of FRBs are still mystery. The detection of FRB 200428 \citep{Bochenek2020,CHIME/FRB2020} in association with a hard X-ray burst \citep{Mereghetti20,CKLi21,konus,AGILE} suggests that at least some FRBs are produced by magnetars at extragalactic distances. Within the FRB context, maser and antenna mechanisms can operate either inside or outside the magnetosphere of magnetars \citep{Katz2018,Lu&Kumar2018,Zhang2020,Lyubarsky2021,Zhang2023}. Pulsar-like models within the magnetosphere invoke two possible radiation mechanisms, i.e. coherent curvature radiation \citep[e.g.][]{Kumar2017,Yang&zhang2018,Wadiasingh19,Kumar&Bosnjak2020,Lu20} and inverse Compton scattering by charged bunches \citep[e.g.][]{Zhang22,QZK,Qu&Zhang2024}. GRB-like models outside the magnetosphere invoke synchrotron radiation in relativistic shocks \citep[e.g.][]{Lyubarsky2014,Beloborodov2017,Beloborodov2020,Plotnikov&Sironi2019,Metzger2019,Margalit2020}.

\begin{figure*}
    \includegraphics[width=18 cm,height=8 cm]{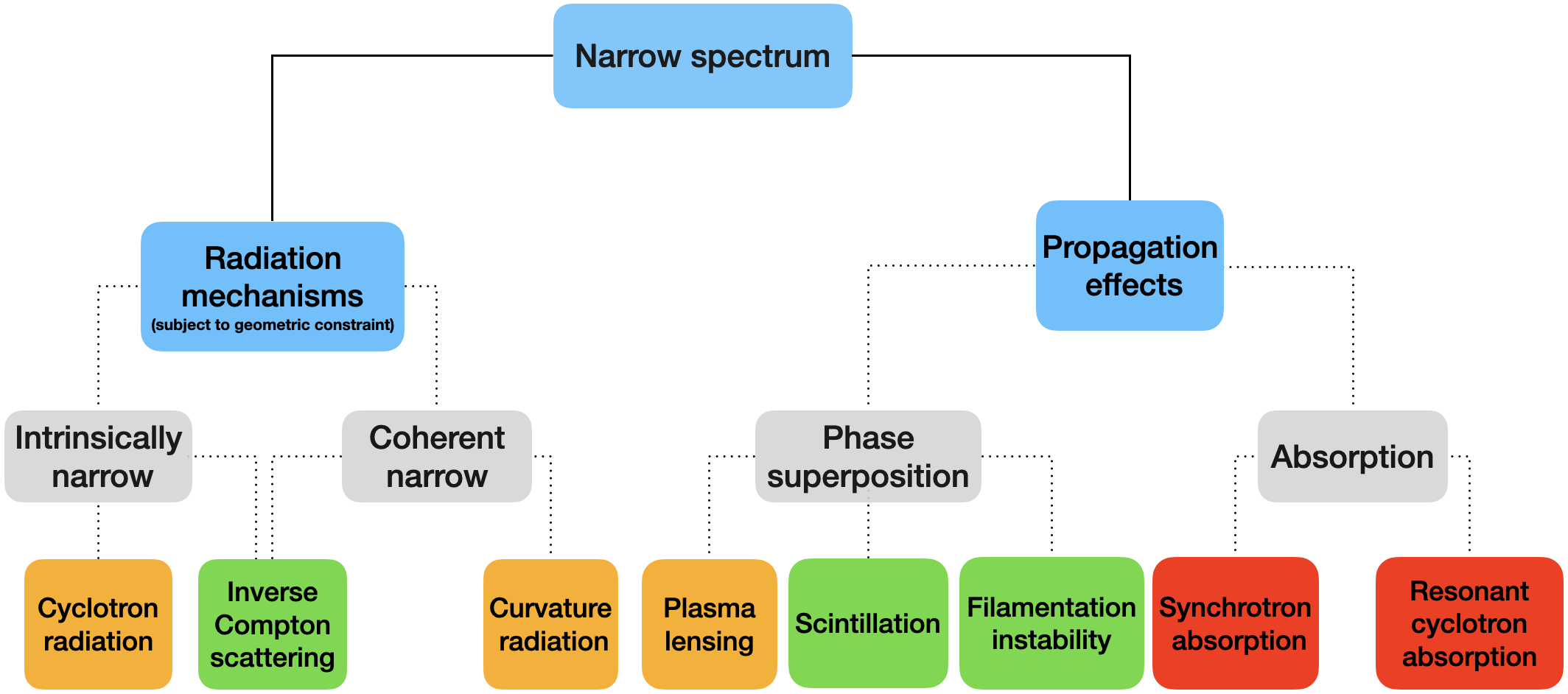}
    \caption{The physical processes to generate narrow spectrum of FRBs discussed in this paper. Two possible ways are considered: intrinsic radiation mechanisms (intrinsically narrow and coherent narrow) and propagation effects (phase superposition and absorption). The favored processes are marked as green. The possible processes under certain physical conditions are marked as orange. The disfavored processes are marked as red.}
    \label{fig:cladogram}
\end{figure*}

The FRBs population is observationally divided into repeating and non-repeating sources. There exists some evidence that the bursts from repeating sources have different properties from the non-repeating FRBs. One important feature is the frequency spectrum, which carries important information about intrinsic radiation mechanisms and propagation effects. Current observational data show interesting but puzzling features regarding narrow spectrum, which we summarize below:
\begin{itemize}
\item The first CHIME/FRB catalog including 62 bursts from 18 repeating FRBs and 474 non-repeating FRBs suggested that repeating FRBs typically have a narrower frequency bandwidth than non-repeater FRBs \citep{Pleunis2021}. Repeating FRBs show Gaussian-like bandwidths (100–200 MHz) in the CHIME band (400-800 MHz) \citep{CHIME2019b}.
\item Not all bursts spectrum can be described by Gaussian-like function. Several non-repeating FRBs extend across the full CHIME bandwidth and the spectrum can be fitted by power-law function \citep{CHIME2019a}.
\item One extremely narrow spectrum burst fitted by a Gaussian function with $\Delta\nu/\nu_0\sim0.05$ (Full-width at half-maximum (FWHM)) from FRB 20190711A which was detected by Parkes radio telescope using the Ultra-wideband Low (UWL) receiver system from 0.7-4.0 GHz \citep{PK2021narrow}, and the integrated S/N can be estimated roughly to be 5. This burst has a bandwidth of 65 MHz, i.e. the spectrum is extremely narrow. 
\item More than 600 bursts from the repeating FRB 20201124A have been detected by Five-hundred-meter Aperture Spherical radio Telescope (FAST) from 1.0 to 1.5 GHz and essentially all bursts have narrow bandwidths \citep{ZhouDJ2022}. The distribution of emission peak frequency has two peaks at 1091.9 MHz and 1327.9 MHz fitted by Gaussian functions. The distribution of bandwidth (defined as the full frequency width at the 10\% of emission peak (FWTM)) has one peak at $\sim 277$ MHz. 
\item Repeating FRB 20220912A including 1076 bursts was detected by FAST and most bursts with extremely narrow spectrum \citep{ZhangYK2023}. The spectra of all bursts of FRB 20220912A are fitted by Gaussian function and the ratio $\Delta\nu/\nu_0$ (FWHM) is centered at 0.1-0.2 implying that the radiation mechanism must account for a narrow spectrum.
\end{itemize}

We use the ratio of $\Delta\nu$ (FWHM) to the central frequency $\nu_0$ to describe the narrowness of an FRB spectrum; a narrow spectrum corresponds to $\Delta\nu/\nu_0<1$. A radiation mechanism may or may not produce narrow spectra, and various processes operating between the source of the FRB waves and the observer might modify the spectra and make them narrower than they were at the source. Figure \ref{fig:cladogram} summarizes the relevant processes discussed in this paper. The intrinsic spectral properties of some of the leading candidate radiation mechanisms for FRBs are discussed in the appendices.

Here we summarize the physical processes that could modify the spectrum of radio waves as they travel from the source to the observer.

\begin{itemize}
\item Plasma lensing: Radio waves traveling through plasma suffer refraction, and that can cause high magnification over a small bandwidth thus result in a narrow observed spetra. This possibility is investigated in \S\ref{sec:plasmalensing}.

\item Scintillation: When an inhomogeneous plasma screen exists between the FRB source and the observer, the wave will travel through different paths and produce interference fringes at the observed plane. The resulting scintillation could modify a broad bandwidth burst to be narrow observed spectrum. The probability of this process is investigated in Sect. \ref{sec:scintillation}.

\item Filamentation instability: FRB waves are likely to be fragmented in the longitudinal and transverse directions as a result of wave-particle interaction as they propagate through the magnetar wind \citep{Sobacchi2021,Sobacchi2022,Sobacchi2023}. This fragmentation might convert a broadband radio pulse to develop a narrow bandwidth structure much like scintillation. We will investigate this in \S \ref{sec:filamentation}. 

\item Absorption process: Another possibility is that some absorption processes (e.g. synchrotron absorption) may diminish low-frequency emission and make the observed spectra narrower. We study synchrotron absorption in Appendix \ref{Appendix:absorption} and conclude that the absorption effects are unlikely responsible for the narrow spectra of FRBs.
\end{itemize}

An FRB radiation mechanisms must explain not only the typical isotropic luminosity $L_{\rm frb}\sim10^{38}-10^{42} \ {\rm erg \ s^{-1}}$ and brightness temperature $T_b\sim10^{36}$ K (see a discussion of energy arguments in \citealt{Lu&Kumar2018}), the high linear and non-zero circular polarization (see a discussion of polarization in \citealt{Qu&Zhang2023}), but also the extremely narrow spectral widths, $\Delta\nu/\nu_0\ll1$, for some FRBs (see a discussion in \citealt{Yang2023}), which will also be investigated in this paper. 

In this paper, we investigate a variety of propagation effects for FRB radio waves that have the potential to explain the narrow spectrum properties of some FRBs.
This paper is organized as follows. In section \ref{sec:general}, we discuss general constraints on the observed spectral narrowness for radiation models that operate inside and others that lie outside the magnetar magnetosphere. 
In section \ref{sec:propagation}, we discuss two possible propagation effects to convert a broad bandwidth spectrum into a narrow observed spectrum. These include plasma lensing and scintillation. 
The probability that scintillation might be able to explain the narrow spectra of three FRBs is also discussed. 
The main results of this investigation are discussed in section \ref{sec:conclusion}. 
A few radiation mechanisms suggested for FRBs, and the effect of the filamentation instability on spectral bandwidth are discussed in Appendices. 
Throughout the paper, the convention $Q=10^{n}Q_n$ in cgs units is adopted.

\begin{figure*}
    \includegraphics[width=16 cm,height=8 cm]{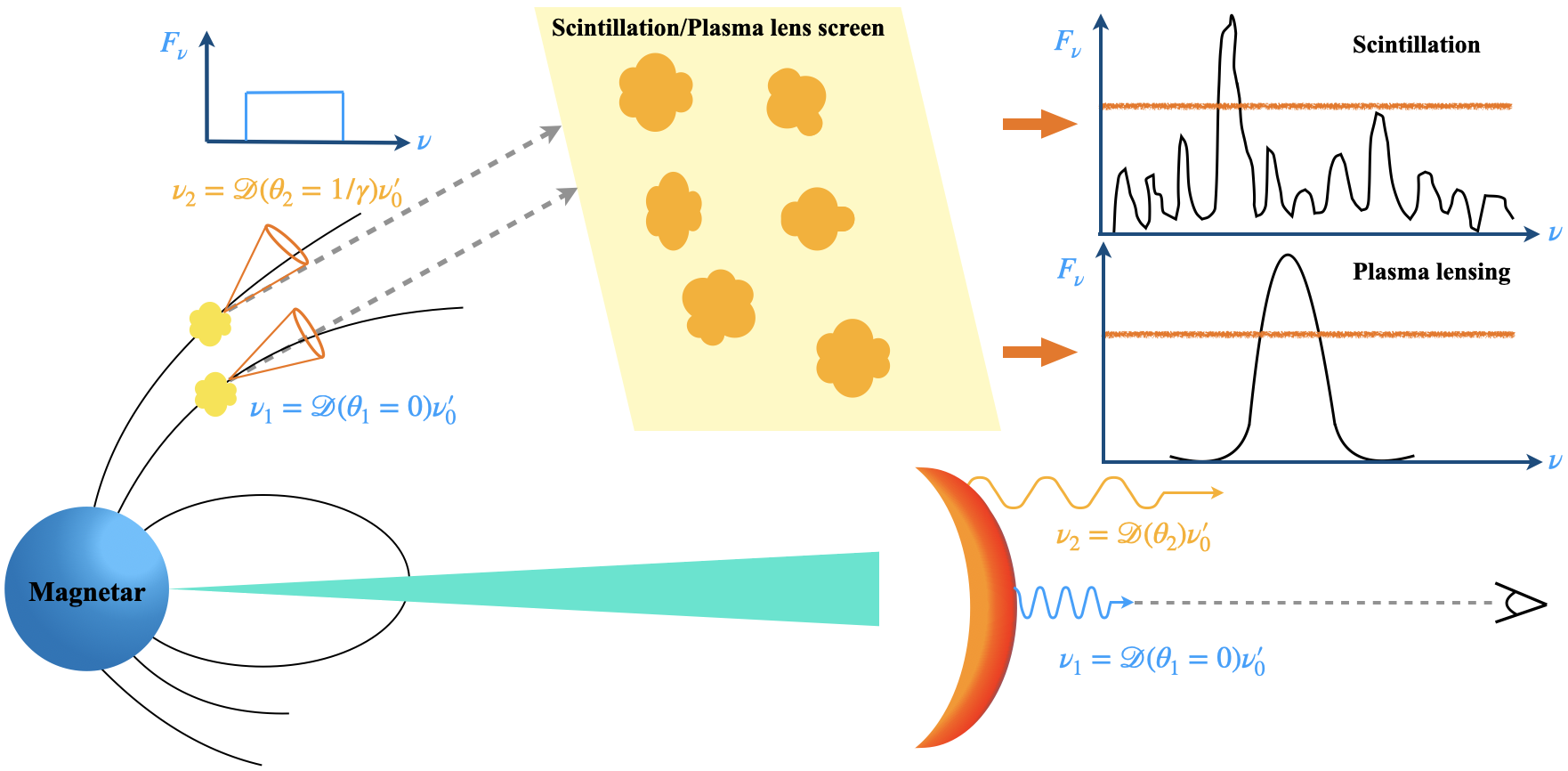}
    \caption{A cartoon figure for emission regions inside and outside the magnetosphere. Inside the magnetosphere, the two yellow clumps move relativistically with a half opening angle of the radiation $\sim 1/\gamma$. Outside the magnetosphere, a spherical thin shell moves radially and photons arrives later with lower frequencies when $\theta_2>\theta_1$ due to the high latitude effect. Only the magnetospheric models radiating FRBs will be strongly influenced by the scintillation/Plasma lens screen and the observed spectrum could be narrow due to scintillation or plasma lensing effect \citep{Kumar2023}.}
    \label{fig:general constraint}
\end{figure*}

\section{A generic constraint}\label{sec:general}

In this section, we provide general constraint on the observed FRB spectrum that is independent of the radiation mechanism. { \cite{Beniamini&Kumar2020} have discussed the radiation properties from a spherical shell including the evolution of peak frequency and flux in time. }

In order to produce the narrowest possible observed spectral bandwidth, we assume that the radiation spectrum is monochromatic in the source comoving frame. 
In the following, 
we consider a thin shell moving towards the observer with the bulk Lorentz factor $\gamma$.
Let us consider a case where the specific intensity and radiation frequency in the comoving frame evolve with time as
\begin{equation}\label{eq:intensity}
I'(\nu',t')=I_0'\,\delta[\nu'-\nu_p'(t')]\left(\frac{t'}{t_0'}\right)^{-\alpha_t},
\end{equation}
and
\begin{equation}
\nu_p'(t')=\nu_0'\left(\frac{t'}{t_0'}\right)^{-\alpha_\nu},
 \label{nupp}
\end{equation}
for $t'>t_0'$. The observed frequency and time can be written in terms of the Doppler factor $\cal D$ as follows
\begin{equation}
\nu={\cal D}\nu', \ \  t_{\rm obs}=\frac{t'}{\cal D}, \quad{\rm where}\quad {\cal D}=\frac{1}{\gamma(1-\beta\cos\theta)},
\end{equation}
$t'$ is the time when a photon is emitted in the source comoving frame, and $\theta$ is the angle between the velocity vector of the source and the line of sight to the observer.
The bandwidth of observed flux, $\delta\nu$, from angle $\theta$ to $\theta+\delta\theta$ at a fixed observer time $t_{\rm obs}$ is
\begin{equation}
\frac{\delta\nu}{\nu}=\frac{d\log{\cal D}}{d\theta}\delta\theta+\frac{d\log\nu_p'}{d\theta}\delta\theta.
\end{equation}
Fixed $t_{\rm obs}$ means that $d\ln t_{\rm obs}/d\theta=0\Rightarrow dt'/(t'd\theta)+\beta\sin\theta/(1-\beta\cos\theta)=0$. Combining these two equations we find
\begin{equation}\label{eq:delta nu/nu}
\frac{\delta\nu}{\nu}=-\frac{(1-\alpha_\nu)\beta\sin\theta\,\delta\theta}{1-\beta\cos\theta} = -(1-\alpha_\nu){\cal D}\gamma\beta\sin\theta\,\delta\theta.
\end{equation}
The frequency integrated contribution to the observed flux that is produced between angles $\theta$ and $\theta+\delta\theta$ is
\begin{equation}
\delta f= \delta\Omega_{\rm obs}\int d\nu\, I_\nu = 2\pi\left(\frac{R}{d}\right)^2\theta\delta\theta\left(\frac{t'}{t_0'}\right)^{-\alpha_t}{\cal D}^4I_0',
 \label{del-f}
\end{equation}
where $\delta\Omega_{\rm obs}=2\pi\delta\cos\theta_{\rm obs}\approx 2\pi \theta_{\rm obs} d\theta_{\rm obs}$, $\theta_{\rm obs} = R\theta/d$, $R$ is the distance of the source from the origin, and $d$ is the distance between the source and observer;
we used Lorentz transformations $I_\nu = I'_{\nu'} {\cal D}^3$, and $\nu = {\cal D}\nu'$ in deriving (\ref{del-f}).
We can eliminate $\theta\delta\theta$ in the expression for $\delta f$ using Eq (\ref{eq:delta nu/nu}), and the definition that $f_\nu=\delta f/\delta\nu$:
\begin{equation}\label{eq:flux}
\begin{aligned}
f_\nu&=\frac{2\pi I_0'}{(1-\alpha_\nu)\gamma}\left(\frac{R}{d}\right)^2\frac{{\cal D}^3}{\nu}\left(\frac{t'}{t_0'}\right)^{-\alpha_t}\\
&=\frac{2\pi I_0'}{(1-\alpha_\nu)\gamma}\left(\frac{R}{d}\right)^2\frac{{\cal D}^{3-\alpha_t}}{\nu}\left(\frac{t_{\rm obs}}{t_0'}\right)^{-\alpha_t}.
\end{aligned}
\end{equation}
Since 
\begin{equation}
t'={\cal D}t_{\rm obs} \ \ {\rm and} \  \ \nu={\cal D}\nu_p'={\cal D}\nu_0'\left(\frac{t'}{t_0'}\right)^{-\alpha_\nu},
\end{equation}
we obtain
\begin{equation}
\nu={\cal D}^{1-\alpha_\nu}\nu_0'\left(\frac{t_{\rm obs}}{t_0'}\right)^{-\alpha_\nu} \ \Rightarrow \ {\cal D}=\left(\frac{\nu}{\nu_0'}\right)^{\frac{1}{1-\alpha_\nu}}\left(\frac{t_{\rm obs}}{t_0'}\right)^{\frac{\alpha_\nu}{1-\alpha_\nu}}
\label{nu-obs3}
\end{equation}
Substituting the above expression for the Doppler factor into Eq.(\ref{eq:flux}) we find
\begin{equation}\label{eq:f_nu(t_obs)}
f_\nu(t_{\rm obs})=\frac{2\pi I_0'}{(1-\alpha_\nu)\gamma\nu_0'}\left(\frac{R}{d}\right)^2\left(\frac{t_{\rm obs}}{t_0'}\right)^{-\alpha_t+\frac{(3-\alpha_t)\alpha_\nu}{1-\alpha_\nu}}\left(\frac{\nu}{\nu_0'}\right)^{\frac{3-\alpha_t}{1-\alpha_\nu}-1}.
\end{equation}

There is a maximum frequency for the radiation received by an observer at a fixed $t_{\rm obs}$, $\nu_{\rm max}$, which corresponds to ${\cal D}(\theta=0)\equiv {\cal D}_{\rm max}=2\gamma$, and a minimum frequency ($\nu_{\rm min}$) that corresponds to ${\cal D}(\theta=\theta_{\rm max})\equiv {\cal D}_{\rm min}\approx2\gamma/(1+\gamma^2\theta_{\rm max}^2)={\cal D}_{\rm max}/(1+\gamma^2\theta_{\rm max}^2)$. We write down the explicit expressions for $\nu_{\rm max}$ \& $\nu_{\rm min}$ using (\ref{nu-obs3}):
\begin{equation}
\nu_{\rm max}=(2\gamma)^{1-\alpha_\nu}\nu_0'\left(\frac{t_{\rm obs}}{t_0'}\right)^{-\alpha_\nu}=\nu_{0,\rm obs}\left(\frac{t_{\rm obs}}{t_{0,\rm obs}}\right)^{-\alpha_\nu},
\end{equation}
and
\begin{equation}
\nu_{\rm min}=\nu_{\rm max}(1+\gamma^2\theta_{\rm max}^2)^{-1+\alpha_\nu} \quad{\rm for} \ \alpha_\nu<1,
\end{equation}
where $\nu_{0,\rm obs}=2\gamma\nu_0'$ and $t_{0,\rm obs}=t_0'/(2\gamma)$. These results are valid only for $t_{\rm obs} > t_{\rm 0,obs}$.

It is convenient to rewrite Eq.(\ref{eq:f_nu(t_obs)}) in terms of $t_{0,\rm obs}$ and $\nu_{0,\rm obs}$ -- 
\begin{equation}\label{eq:observed flux}
f_\nu(t_{\rm obs})=\eta\times\left(\frac{t_{\rm obs}}{t_{0,\rm obs}}\right)^{\frac{3\alpha_\nu-\alpha_t}{1-\alpha_\nu}}\left(\frac{\nu}{\nu_{0,\rm obs}}\right)^{\frac{2-\alpha_t+\alpha_\nu}{1-\alpha_\nu}},
\end{equation}
where 
\begin{equation}
\eta=\frac{4\pi I_0'(2\gamma)^2}{(1-\alpha_\nu)\nu_{0,\rm obs}}\left(\frac{R}{d}\right)^2,
\end{equation}
and $\nu$ should be between $\nu_{\rm min}$ and $\nu_{\rm max}$. We can see from Eq.(\ref{eq:observed flux}) that the burst duration is nearly independent of frequency. Therefore, the frequency dependence of the observed fluence is roughly the same as the specific flux.
The time dependence of the flux at $\nu_{\rm max}$ is
\begin{equation}
f_{\nu_{\rm max}}(t_{\rm obs})=\eta\left[\frac{t_{\rm obs}}{t_{0,\rm obs}}\right]^{\alpha_\nu-\alpha_t}.
\end{equation}
For $\alpha_t=0$, $f_{\nu_{\rm max}}(t_{\rm obs})\propto t_{\rm obs}^{\alpha_\nu}$, i.e. the specific flux at $\nu_{\rm max}$ increases with time for $\alpha_\nu > 0$.
However, by assumption, the bolometric flux is $\propto t'^{-\alpha_t}\propto t_{\rm obs}^{-\alpha_t}$, which is independent of time when $\alpha_t=0$. 
Since, the bolometric flux in the observer frame is $\sim\nu_{\rm max} f_{\nu_{\rm max}}\propto t_{\rm obs}^{-\alpha_{\nu}}t_{\rm obs}^{\alpha_\nu}$, it is also time independent as it should be.
The spectrum at a fixed observer time is proportional to $\nu^2$ for $\alpha_t=0$ \& $\alpha_\nu=0$ (Eq.\ref{eq:observed flux}). 
Thus, the observed specific fluence in this case is: 
\begin{equation}
F(\nu)= F_{\nu_{\rm max}} (\nu/\nu_{\rm max})^2,
  \label{high-lat-spect}
\end{equation}
which is 50\% smaller than the peak value at $\nu=\nu_{\rm max}/\sqrt{2}$, i.e. the half-width at half maximum of the observed signal is $\sim \nu_{\rm max}/\sqrt{2}$. This sets the minimum bandwidth of the observed signal, even though the FRB source is assumed to be monochromatic in its rest frame. However, this result is subject to the condition that the angular size of the source is larger than $\gamma^{-1}$. We now discuss the implications of this result for the two proposed classes of FRB mechanisms: sources located inside the magnetosphere of a NS, and relativistically moving sources producing coherent radio pulses outside the light-cylinder.

For FRB sources outside the magnetosphere, the lateral size of the source region is expected to be roughly of order of its distance from the compact object that produced the relativistic outflow, which in turn generated the coherent radio waves. Consequently, the angular size of the source region is approximately one radian, significantly larger than the Doppler beaming angle of $\gamma^{-1}$. In this scenario, the half-width of the observed spectrum is given by $\nu_{\rm max}(1 -1/\sqrt{2})$ (see eq. \ref{high-lat-spect}), assuming the intrinsic spectrum in the source's comoving frame is monochromatic. And the FWHM of the observed spectrum is
\begin{equation}
\frac{\Delta\nu}{\nu_{\rm max}}=2-\frac{2}{\sqrt{2}}\simeq0.58.
\end{equation}

For FRB sources inside the magnetosphere, the typical emission radius might be $r_{\text{em}}\sim10^8$ cm, the Lorentz factor of the plasma is $\gamma\sim100$, and the transverse size of the source could be on the order of the coherence length dictated by the causality argument, i.e., $\gamma c/\nu \sim 3\times10^{3}\nu_9^{-1}$ cm. Hence, the angular source size is $\theta_s\sim 3\times10^{-5}\nu_9^{-1}/r_{\text{em},8}$ rad, which is much smaller than $\gamma^{-1}$ as long as $\gamma \ll 2\times10^3$. Therefore, the observed bandwidth, $\Delta\nu/\nu_{\rm max}\sim (\theta_s\gamma)^2$, can be very small for a monochromatic source inside the magnetosphere.

{
It should be pointed out that the re-absorption process of the emitted photons can only affect photons at angles greater than $1/\gamma$, which will not influence the intrinsic FWHM ratio $\Delta\nu/\nu_{\rm max}\simeq0.58$ due to the high latitude effect. 
The radiation produced from the synchrotron maser model could be highly beaming if the magnetization is much greater than unity. However, the efficiency will drop quickly and only 100\% linear polarized waves can be produced, e.g. \cite{Beniamini&Kumar2020} }

To summarize the main results of this section, we find that the observed spectrum cannot be narrower than $\Delta\nu/\nu_{\rm max}\simeq0.58$ for an FRB source outside the NS magnetosphere. This limitation arises due to the spectral broadening effect caused by high-latitude emission when the angular size of the relativistic shell producing the radio pulse exceeds the Doppler beaming angle. The lower limit on $\Delta\nu/\nu_{\rm max}$ of 0.58 is derived under the assumption of a monochromatic source in its comoving frame. Additionally, if the comoving frame frequency changes with time, as it must for a maser operating in shocked plasma, that also contributes to the broadening of the observed spectrum. In reality, the intrinsic spectrum is unlikely to be monochromatic, resulting in the full width at half maximum (FWHM) of the observed spectrum being broader than $0.58\,\nu_{\rm max}$. For an FRB source inside the magnetosphere, the spectral bandwidth can be arbitrarily narrow, as the angular size of the source can be much smaller than the Doppler beaming angle.

\section{Propagation effects}\label{sec:propagation}
There are three propagation effects that will be discussed in this section. The first of which is plasma lensing which could magnify FRB waves in a narrow frequency band and thereby could be responsible for the observed small bandwidth for several FRBs. The second effect is the scintillation of FRB waves due to the fluctuation of plasma density along the propagation path. And the third effect is the break-up of FRB pulse into filaments as a result of its interaction with plasma resulting in the index of refraction varying with wave amplitude and plasma density. 

\subsection{Plasma lensing and its effect on possibly making the spectrum narrow}\label{sec:plasmalensing}

The scenario explored in this subsection, pertaining to a narrow bandwidth of the observed spectrum, involves high-magnification plasma lensing. We contemplate the existence of a plasma lens between the source and the observer, which greatly amplifies the flux at the peak of the observed spectrum. 
The frequency-dependent nature of this magnification could potentially account for the observed narrow spectrum, even when the intrinsic spectrum is broad band. We calculate the frequency dependence of the magnification factor, and the observed bandwidth.

The wave amplitude at the observer location when there is a plasma screen between the source and the observer is given by the Fresnel-Kirchoff integral over the plasma screen surface, e.g. \citep{jow2021,Feldbrugge2019}
\begin{equation}\label{eq:Ai}
A_i(\nu) \propto \int d^2{\vec \theta} \exp\left[ \frac{ i 2\pi (\vec\theta-\vec\theta_s)^2}{\theta_F^2}+ i\phi_p(\vec\theta,\omega)\right],
\end{equation}
where $\vec\theta_s$ is the angular location of the source in the sky wrt observer-lens line, $\vec\theta$ is
the angular position of a point in the plasma lens plane,
\begin{equation}\label{fresnel}
\theta_F=\left(\frac{2\lambda d_{\rm SL}}{d_{\rm LO}d_{\rm SO}}\right)^{1/2}
\end{equation}
is the Fresnel angle, $d_{\rm SL}$ is the distance from the FRB source to the lens, $d_{\rm LO}$ from the lens to the observer, and $d_{\rm SO}$ from the FRB source to the observer. 
The Fresnel angle is defined as the angular position on the lens plane, such that a photon from the source passing through this point and arriving at the observer undergoes an additional geometric phase shift of $2\pi$ compared to a straight-line trajectory from the source to the observer,
and  $\phi_p(\vec\theta,\omega)$ is the phase shift suffered by the wave as it crosses the plasma screen at
the angular location $\vec\theta$. The phase shift $\phi_p$ is given by
\begin{equation}
\phi_p(\theta,\omega)=-\omega\int \frac{dz}{c}\frac{2\pi q^2 n_e(\vec\theta, z)}{m_e\omega^2} \equiv -\omega^{-1} \xi N_e(\vec\theta),
\end{equation}
where $z$ is the coordinate perpendicular to the plasma lens plane,
\begin{equation}
\xi\equiv\frac{2\pi q^2}{m_e c}=0.053\,{\rm cm^2\,s^{-1}} \ {\rm and} \  N_e(\vec\theta)\equiv\int dz n_e(\vec\theta,z)
\end{equation}
is the electron column density of the lens at the angular position $\vec\theta$.
The angular location of an image is given by the extremum of the phase function
\begin{equation}\label{phase-fun}
\Phi \equiv \frac{2\pi (\vec\theta - \vec\theta_s)^2}{\theta_F^2}+\phi_p(\vec\theta,\omega),
\end{equation}
or explicitly,
\begin{equation}\label{image-loc}
\frac{\partial \Phi}{\partial \theta_\mu} \equiv \partial_\mu \Phi=0 \ {\rm or} \
\frac{4\pi\theta_{i \mu}}{\theta_F^2}-\omega^{-1} \xi \partial_\mu N_e(\vec\theta_i)=0,
\end{equation}
where we took the source location $\vec\theta_s=0$ by shifting the origin of the coordinate system without loss of generality. We see from the above equation that the angular position of the image in the sky is dependent on frequency.

The amplification of the flux due to the plasma lens is given by
\begin{equation}
{\cal M} = \frac{16 \pi^2}{ \theta_F^4 {\rm det}\left[ \partial_\mu \partial_\nu \Phi \right] }.
\label{mag}
\end{equation}
We can write the elements of the $2\times2$ `amplification' matrix by using Eq. (\ref{phase-fun}) as
\begin{equation}
\partial_\mu \partial_\nu \Phi = 4\pi \theta_F^{-2} \delta_{\mu\nu} - \omega^{-1} \xi \partial_\mu\partial_\nu N_e(\vec\theta),
\end{equation}
where $\delta_{\mu\nu} = 1$ for $\mu=\nu$ and zero for $\mu\not=\nu$, meaning it is the standard Kronecker delta function.
Since $\partial_\mu \partial_\nu \Phi $ is a $2\times2$ symmetric matrix, it can be diagonalized by a coordinate rotation. It is convenient to write the matrix as
\begin{gather}
\partial_\mu\partial_\nu N_e(\vec\theta) =
\begin{pmatrix}
a_1  & 0 \\
0   & a_2
\end{pmatrix}
 \ {\rm \&} \  \partial_\mu \partial_\nu \Phi = {4\pi\over \theta_F^2}
\begin{pmatrix}
b_1^{-2}  &  0 \\
0  &  b_2^{-2}
\end{pmatrix}
\label{mag-mat}
\end{gather}
where
\begin{equation}
  b_i \equiv {1\over \sqrt{1 - \psi a_i}}, \; \psi \equiv {\theta_F^2 \xi\over 4\pi \omega} = \psi_0 \left( {\omega_0\over\omega}\right)^2, \;
      \psi_0 \equiv {\theta_F^2 \omega\xi\over 4\pi\omega_0^2},  
\end{equation}
and $\omega_0$ is the frequency where the magnification for the plasma lens peaks. We note that
$\psi_0$ is independent of wave frequency since $\theta_F^2 \propto \omega^{-1}$. The shape of the region in the lens plane from which all waves arrive at the observer with a relative phase difference of $<\pi$, and therefore their amplitudes add constructively, can be approximated as an ellipse when cubic and higher-order terms in angle are neglected. Waves coming from outside this ellipse contribute little to the observed flux. The angular sizes of the two axes of this ellipse are $\theta_F b_1$ and $\theta_F b_2$. For a circular lens, $b_1=b_2$, and for a highly elongated lens either $b_1 \gg b_2\sim 1$ or $b_2 \gg b_1\sim 1$. The magnification-factor of the plasma lens can be written explicitly using Eqs. (\ref{mag}) \& (\ref{mag-mat}) as
\begin{equation}
      {\cal M} = \prod_{i=1}^2b_i^2 = \prod_{i=1}^2 \left[ 1 - \psi_0 a_i (\omega_0/\omega)^2\right]^{-1}
         \equiv {\cal M}_1 {\cal M}_2
  \label{mag2}
\end{equation}
where
\begin{equation}
    {\cal M}_i(\omega) \equiv  b_i^2 \equiv \left[ 1 - \psi_0 a_i(\omega) \omega_0^2/\omega^2\right]^{-1}
 \label{mag3}
\end{equation}
Thus the magnification is proportional to the square of the area of the ellipse, which makes good sense as we can see from equation (\ref{eq:Ai}) that the wave amplitude is proportional to the area of the ellipse as the phase factor in that part of the lens is constant and $\exp(i\Phi)\sim 1$.

The magnification can only be achieved when the angular size of the source is smaller than $\theta_F/{\cal M}^{1/4}$ for a nearly circular plasma lens, and $\theta_F/{\cal M}^{1/2}$ for a highly elongated elliptical lens\footnote{See e.g. the ``Methods" section of \cite{Main2018} for a simple derivation of these results. Waves from a source of an angular size larger than $\theta_F$ that pass through the plasma lens do not interfere constructively, as is the case for a smaller source. As a result of this cancellation, the magnification is reduced to order unity.}. As an example, for a plasma lens at a distance $R$ from the source, no magnification is produced at wavelength $\lambda$ when the source size is $\gg (\lambda R)^{1/2} \sim 10^9 R_{17}^{1/2} \nu^{-1/2}$cm. Thus, we don't expect FRB flux to be magnified if it is generated in an external shock as the expected source size in that case is $\geq 10^{10}$cm.

We provide an order of magnitude estimate for electron density required for a plasma lens to magnify FRB flux by a large factor. The requirement for strong amplification is that the phase shift suffered by the wave crossing the plasma should almost cancel the geometric phase shift over a region of area much larger than $\theta_F^2$. We know from Eq. (\ref{mag-mat}) that the linear size of the region where this condition is satisfied is $\theta_F \max\{b_1, b_2\}$. For a circular lens, the geometric phase shift across this region is, therefore, $2\pi {\cal M}^{1/2}$.  Thus, we find that the plasma phase shift must satisfy: $\xi \delta N_e/\omega \approx 2\pi {\cal M}^{1/2}$; where $\delta N_e(\vec\theta) \equiv N_e(\vec\theta) - N_e(\vec\theta_I)$ with $\vec\theta_I$ being the center of the ellipse or the image location. Therefore, $\delta N_e \approx 2\pi \omega {\cal M}^{1/2}/\xi = {\rm 10^{12}\, {\cal M}^{1/2}\nu_9\, cm}^{-2}$. Moreover, the functional form of $\delta N_e(\vec\theta)$ should almost exactly trace the angular dependence of the geometric phase-shift so that the two phase-shift terms cancel in the elliptical region and lead to a strong magnification.
We note that the electron column density at $\vec\theta$, $N_e(\vec\theta_I)$, is much larger than the $\delta N_e(\vec\theta)$ estimated above if $\theta_I$, the angle between the source-observer line of sight and the image location, is an arc-second or larger. The phase shift of the wave due to plasma should be such that the wave is refracted by the angle $\theta_I$, i.e. $\theta_I \sim \Phi(\vec\theta)\lambda^2/(d_F^2 \theta_I)$. Or, $N_e(\vec\theta_I)\sim d_F^2 c \theta_I^2/(\xi\lambda^3)\sim 2{\rm x}10^{17} d_{F,10}^2 \theta_{I,-5}\nu_9^3$cm$^{-2}$; where $d_F\equiv (2\lambda d_{SL} d_{LO}/d_{SO})^{1/2}$ is the Fresnel radius, and $\theta_{I,-5}$ is the angular position of the image wrt source-observer line in units of 10$^{-5}$ radian.

The frequency bandwidth for the magnification $\cal M$ can be calculated from the frequency dependence of the phase function which is given by 
\begin{equation}
    \Phi(\theta, \omega) = {2\pi \theta^2\over \theta_F^2(\omega_0)} {\omega\over\omega_0} - {\xi N_e(\vec\theta) \over \omega_0} {\omega_0\over\omega},
    \label{density-needed}
\end{equation}
where we have taken $\vec\theta_s = 0$ by exercising the freedom to choose the origin of the coordinate system, and we have factored out the frequency dependence of each term in the above equation explicitly. At frequency $\omega_0$, the two terms in the above equations are equal to within $\pi/2$ over the entire area of the ellipse from where waves are focused at the observer. The center of the ellipse is at $\vec\theta_I$, which is the location of the image of the source. The maximum geometric phase difference between the center of the ellipse and the end of one of its two principal axes is
\begin{equation}
    \delta\Phi_G = 2\pi {\left[ \vec\theta_I + \theta_F b_\mu \, \hat\mu\right]^2 - \theta_I^2 \over \theta_F^2} = 2\pi\left(b_\mu^2 + 2\theta_{I\mu} b_\mu/\theta_F\right),
\end{equation}
where $\theta_{I\mu}$ is the component of $\vec\theta_I$ along principal axis $\mu$ which is oriented along the unit vector $\hat\mu$. 
As mentioned already, the phase difference suffered by the wave as it goes through the plasma within the ellipse has a magnitude that is almost exactly the same as $\delta\Phi_G$ but of the opposite sign at frequency $\omega_0$. However, at a different frequency $\omega_0+\delta\omega$, the difference between the two phases is $\sim (2\delta\omega/\omega_0) \delta\Phi_G \sim 4\pi b_\mu^2 (\delta\omega/\omega_0)$, when $|\theta_{I\mu}| \ll \theta_F b_\mu$.  When this phase difference is $\gg\pi$, waves arriving at the observer from different parts of the ellipse cancel, and the magnification is reduced. Thus, the bandwidth for strong lensing is
\begin{equation}
    {\delta\omega_{\rm lens}\over\omega_0} \sim \min\left\{ {1\over b_1^2}, \, {1\over b_2^2} \right\}.
\end{equation}

For a circular lens, $b_1 \sim b_2$, and the magnification ${\cal M} \approx b_1^4$ (Eq. (\ref{mag2})), and therefore the bandwidth can be written as
\begin{equation}
\frac{\delta\omega_{\rm lens}}{\omega}\sim\frac{1}{{\cal M}^{1/2}}.
\end{equation}
Whereas for a highly elongated lens, ${\cal M} \sim \max\{b_1^2, b_2^2\}$, therefore,
\begin{equation}
    {\delta\omega_{\rm lens}\over\omega} \sim {1\over \cal M}.
\end{equation}

The calculation above for the bandwidth of the lens only showed that $\cal M$ should change when the frequency changes by $\delta\omega_{\rm lens}$. It does not consider the possibility that the location of the image changes with frequency, and so do the axes of the ellipse that focus the wave at the observer location. When these changes are properly included in the calculation it might turn out that the magnification does not peak at $\omega_0$. We are interested in this work in determining whether magnification by a plasma lens could be responsible for the narrow bandwidth spectra seen for some FRBs. For this scenario to work, the magnification should peak at some frequency within the observational band and the magnification should decrease rapidly over a small frequency interval about this peak so that flux-limited observations find the spectrum to be narrow-band. For this purpose, we expand $\cal M$ in Taylor series about its peak at $\omega_0$. The linear term in the Taylor series vanishes since
the magnification peaks at $\omega_0$. Thus, the series is
\begin{equation}
   \ln{\cal M}(\omega) =  \ln{\cal M}(\omega_0) + \zeta {(\delta\omega)^2\over \omega_0^2} +
        \mbox{ higher order terms}
   \label{mag4}
\end{equation}
where
\begin{equation}
  \begin{split}
   \zeta  & = {d^2 \ln {\cal M}\over d\omega^2} \bigg|_{\omega_0} \omega_0^2 \\
     & = \prod_{i=1}^2 \Bigg\{ ({\cal M}_i - 1)^2  \left[
    \kappa_i - 2\right]^2
     + ({\cal M}_i - 1) \times \\
   & {\hskip 3cm} \left[ 6 - 5 \kappa_i + \kappa_i^2 + \kappa_i
        {d\ln\kappa_i\over d\ln\omega} \right] \Bigg\}
  \end{split}
   \label{zeta}
\end{equation}
and
\begin{equation}
  \kappa_i \equiv {d\ln a_i\over d\ln\omega}.
   \label{ki}
\end{equation}
All quantities in Eqs. (\ref{zeta}) \& (\ref{ki}) are evaluated at $\omega=\omega_0$.
The lens equation (Eq. (\ref{image-loc})) shows that the frequency dependence of image location
follows the relation $d\ln\theta_{i\mu}/d\ln\omega\sim -2$. Therefore, as long as the electron column density in
the plasma lens varies smoothly with $\vec\theta$, we expect $|\kappa_i|\sim$ O(1);
additionally, the dimensionless derivative of the logarithm of a smoothly varying function is
usually on the order of unity. For similar
reasons, we have $|d\ln\kappa_i/d\ln\omega|\sim$ O(1). Therefore, for high magnification events
(${\cal M} \gg 1$) with $|\kappa_i -2|\not\ll 1$ we expect $\zeta \sim {\cal M}^2$, and the bandwidth of the spectrum due to strong plasma lensing is
\begin{equation}
    \delta\omega_{\rm lens} \sim {\omega_0 \over {\cal M}(\omega_0) }.
\end{equation}
However, when $\kappa_i\approx 2$, $\zeta \propto {\cal M}$, and the lensing bandwidth is
\begin{equation}
    \delta\omega_{\rm lens} \sim {\omega_0 \over {\cal M}^{1/2} (\omega_0) }.
\end{equation}

With this rough scaling for $\zeta$,
we can rewrite the frequency dependence of the magnification factor in the neighborhood of
$\omega_0$ using Eq. (\ref{mag4}) as
\begin{equation}
    {\cal M}(\omega) \sim {\cal M}(\omega_0) \exp\left\{ -\left[{\delta\omega
      {\cal M}(\omega_0) \over \omega_0} \right]^2 \right\}.
      \label{plasma-mag4}
\end{equation}
The negative sign in the exponent arises from the fact that the peak of ${\cal M}(\omega)$ occurs at $\omega_0$, and the observed flux decreases as $|\omega-\omega_0|$ increases. The calculation of lens magnification presented here provides a quantifiable measure of the extent to which this mechanism can generate a narrow spectrum.
The observed flux decreases with frequency by a factor $e$ as we move away from the
peak of the spectrum by
\begin{equation}
   \delta\omega_{\rm lens} \sim {\omega_0 \over {\cal M}(\omega_0) }.
\end{equation}

If the narrowness of the observed spectrum is a result of high magnification lensing, the data should be symmetric around the peak with an approximately Gaussian shape. This characteristic can be examined to ascertain whether the narrow spectrum is indeed attributed to strong magnification. Moreover, if the burst is a repeater and most of its outbursts display narrow spectra, it is less likely that the cause is a high-magnification lensing mechanism. This is because the surrounding medium near the FRB source, where the lens is perhaps situated, is likely dominated by the magnetar wind and should change rapidly over short timescales.

The bottom line is that narrow spectrum could be due to highly chromatic plasma lensing with large magnification. 
The electron column density needed for a strong plasma lens is modest, around 10$^{12}$cm$^{-2}$ (see eq. \ref{density-needed}). However, even a small random fluctuation in the density across the Fresnel radius can disrupt the delicate requirement on electron density provide for high magnification plasma lensing, leading to scintillation instead. 
For example, a fluctuation of $10^{-3}$ cm$^{-3} R_{15}^{-1}$ in electron density across the Fresnel radius can render the lens ineffective; $R$ is the distance between the source and the lens. 
This limit on density fluctuation suggests that high magnification plasma lens should be situated close to the FRB source, within a distance on the order of 10$^{15}$cm. Additionally, magnification requires sources with sizes smaller than the Fresnel radius divided by the lens's magnification. 
Therefore, the limit on the source size is typically around 10$^6$cm. This limitation rules out the feasibility of strong lensing as an explanation for narrow spectra when the FRB pulse is produced outside the magnetosphere.

\subsection{Effect of Scintillation on the observed bandwidth}\label{sec:scintillation}

The wavefronts of FRB waves are likely tilted and the rays will be deflected at a characteristic refraction angle $\theta_s$ by the plasma screen, then different waves will interfere with each other. 
As a result, the observed intensity fluctuates with frequency caused by the combination of multiple waves that traveled through different paths. 
This propagation effect, called scintillation, has a non-zero probability to produce a narrow spectrum, i.e., $\Delta\nu/\nu_0<1$, for some bursts with small values of S/N. In this subsection, we investigate the effect of scintillation on the observed bandwidth and quantify the probability of it transforming a broadband source into a narrow observed spectrum. The plasma screen could be located close to the FRB source or midway between the source and observer. Alternatively, it could be in the Milky Way close to the observer.

Let us first consider the plasma screen a distance $D$ from the source or observer, whichever is closer to the screen.
The scintillation frequency bandwidth is given by \citep{LynePulsar,Draine}
\begin{equation}
\delta\nu_{\rm scint}=\frac{8\pi^2m_e^2cL}{e^4D^2\Delta n_e^2}\nu^4\simeq (370 \ {\rm MHz}) \ L_{13}\nu_9^{4}D_{21}^{-2} n_{e,-3}^{-2},
\end{equation}
where $L$ is the size of largest eddies, and $n_e$ is the electron density fluctuation of scale $L$.  
The scintillation bandwidth becomes larger for higher frequency waves, thus if the observed narrow spectrum is due to scintillation, we should see its bandwidth increasing rapidly with frequency.  

The interstellar medium in the Milky Way can also cause FRB waves phase variations via scintillation. The empirical bandwidth for the Milky Way is given by \citep{Cordes&Chatterjee2019}
\begin{equation}
\delta\nu_{\rm scint,MW}\simeq(4 \ {\rm MHz}) \ |\sin{b}|^{6/5}\nu_9^{4.4},
\end{equation}
where $b$ is the latitude. \cite{Kumar2023} discussed the scintillation effect of two plasma screens (one is close to the FRB source at a distance of order 1 pc and another is further away in the host galaxy). In the far-away model for FRBs, where radio emission is generated outside the light-cylinder, scintillation is suppressed because the source size is larger than the diffractive length of the scattering plasma screen. For FRB waves produced inside the magnetosphere, the source size is much smaller, resulting in flux variations due to scintillation with amplitudes of the order of unity.

The probability density function (PDF) of the wave amplitude $A(\nu)$ due to scintillation can be described by Rayleigh distribution \citep{LynePulsar}. Thus, the PDF of the observed flux $f(\nu)\sim A^2(\nu)$ can be written as
\begin{equation}
F[f(\nu)]={\exp [-f(\nu)/f_0(\nu)]\over f_0(\nu)},
\end{equation}
where $f_0$ is the flux in the absence of the scintillation screen. The total probability is 1, i.e. $\int_{0}^{\infty}F[f(\nu)]=1$. The probability for the observed flux to lie between $f_1$ and $f_2$ is
\begin{equation}
\begin{aligned}
P(f_1,f_2)=\int_{f_1}^{f_2}F(f)df&={\rm exp}[-f_1/f_0]-{\rm exp}[-f_2/f_0].
\end{aligned}
\end{equation}
Let us break up the observing band $\nu_1$--$\nu_2$ into $N=(\nu_2-\nu_1)/\delta\nu_{\rm scint}$ channels. To explain an observed narrow spectrum as a result of scintillation where the observed flux is below the detection threshold flux $f_{\rm min} \equiv f_0 \alpha$ over all frequencies except between $\nu_{1a}$ \& $\nu_{2b}$, the observed flux is $f_{\rm obs}=S_N f_{\rm min}=S_N\alpha f_0$.
The probability that the observed flux is equal to or larger than $f_{\rm obs}$ in a frequency band of width $\delta\nu_{\rm scint}$ is
\begin{equation}
P(f_{\rm obs},\infty)=\exp(-S_N\alpha),
\end{equation}
the probability for the flux in another frequency channel of the same bandwidth to be less than $f_{min}=f_0\alpha$ is
\begin{equation}
P(0,f_{\rm min})\simeq 1-\exp(-\alpha).
\end{equation}
and the probability for the flux larger than the detection threshold but smaller than the observed maximum flux can be written as
\begin{equation}
P(f_{\rm min},f_{\rm obs})={\rm exp}(-\alpha)-{\rm exp}(-S_N\alpha).
\end{equation}

Since the parameter $\alpha$ is uncertain, in the following, we discuss two cases: (i) a constant $\alpha$ and a flat spectrum incident on the scintillation screen; (ii) a frequency dependent $\alpha$ and a Gaussian spectrum, i.e. $f_0=A{\rm exp}[-(\nu-\nu_0)^2/2\sigma^2]$, $A$ is the flux amplitude in the absence of scintillation, $\nu_0$ is the central frequency.

For the case (i), the joint probability for non-zero flux observed in the interval ($\nu_{1a}, \nu_{2b}$) and flux below the detection threshold at other frequencies due to scintillation can be calculated as
\begin{equation}
P_{\rm obs} \sim \prod_{i=1}^{n_1} P(0,f_{\rm min}) \prod_{i=1}^{n_2}  p(f_{\rm obs},\infty),
\end{equation}
where
\begin{equation}
n_1 \equiv { \nu_2 - \nu_1 + \nu_{1a} - \nu_{2b} \over \delta\nu_{\rm scint} } \; \& \; n_2 \equiv { \nu_{2b} - \nu_{1a}\over \delta\nu_{\rm scint} }.
\end{equation}
The joint probability can be re-written as
\begin{equation}
P_{\rm obs}(\alpha) \sim \Bigl[P(0,f_{\rm min}) \Bigr]^{n_1} \times \;\Bigl[ P(f_{\rm obs},\infty) \Bigr]^{n_2}.
  \label{obs-p}
\end{equation}

\begin{figure*}
\begin{center}
\begin{tabular}{ll}
\resizebox{95mm}{!}{\includegraphics[]{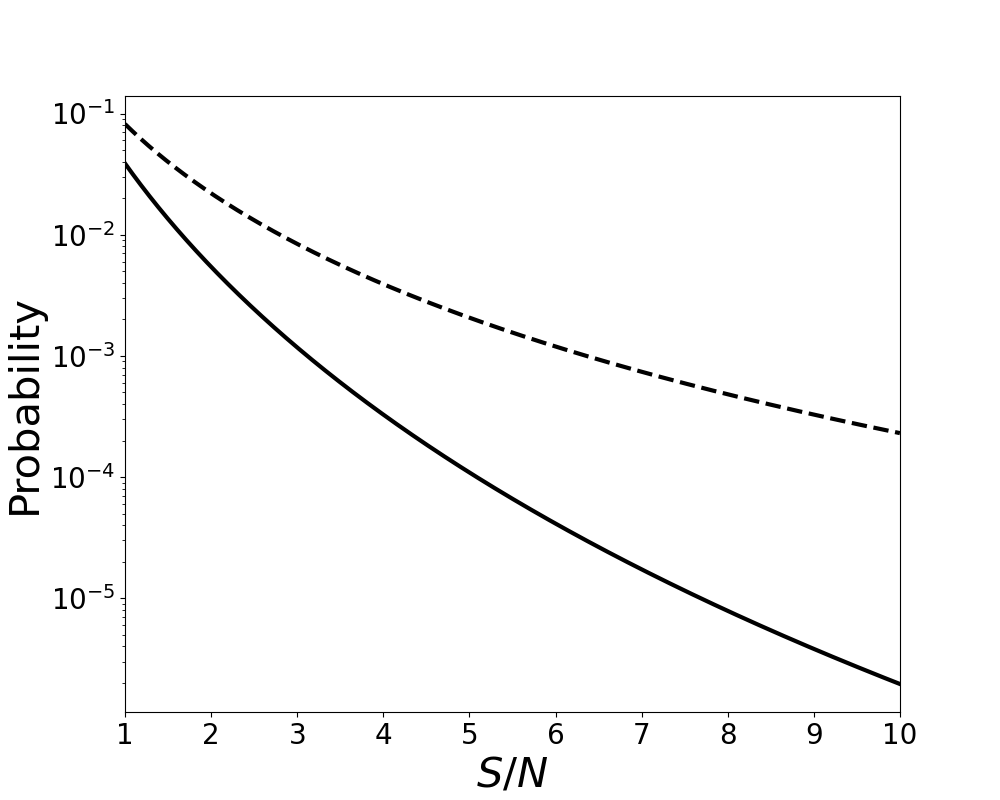}}&
\resizebox{82mm}{!}{\includegraphics[]{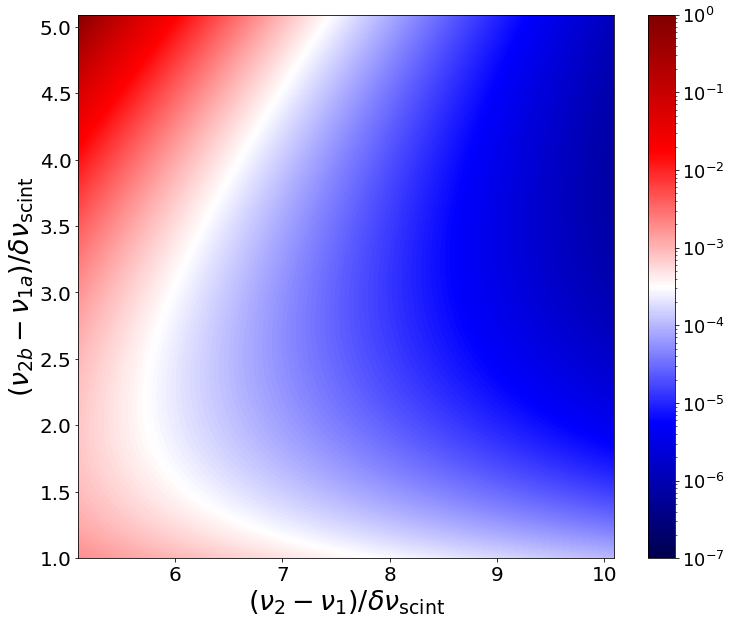}}
\end{tabular}
\caption{Left panel: Probability of scintillation as a function of S/N for different bandwidth ($\nu_2$-$\nu_1$) of detectors. 
The bandwidth of the detector is $10\delta\nu_{\rm scint}$ (black solid line) or $5\delta\nu_{\rm scint}$ (black dashed line); the bandwidth of the observed spectrum is taken to be $\delta\nu_{\rm scint}$, which is the scintillation frequency bandwidth. Right panel: The probability as a function of $(\nu_2-\nu_1)/\delta\nu_{\rm scint}$ and $(\nu_{2b}-\nu_{1a})/\delta\nu_{\rm scint}$; the observation band is $[\nu_1,\nu_2]$ and the flux is below the detection threshold over all frequencies except between $\nu_{1a}$ \& $\nu_{2b}$.  The value of S/N is fixed to 5.}
\label{fig:P_scint}
\end{center}
\end{figure*}

Since we don't know the value of $\alpha$ from observations, we calculate the maximum value of the function $P_{\rm obs}(\alpha)$, which is given by Eq. (\ref{obs-p}). Then the maximum probability of the narrow bandwidth spectrum due to scintillation can be written as
\begin{equation}
P_{\rm obs}(\alpha_{\rm max}) = \left({n_1\over n_1 + n_2 S_N}\right)^{n_1} \times \left({n_2 S_N\over n_1 + n_2 S_N} \right)^{n_2 S_N},
 \label{scint-prob4}
\end{equation}
where 
\begin{equation}
\alpha_{\rm max}=-\ln\left( {n_2 S_N\over n_1 + n_2 S_N} \right).
\end{equation}
For example, let us consider an FRB spectrum that is detected with $S_N=5$, and the signal is detected in a frequency bin of size $\delta\nu_{\rm scint}$, and the rest is consistent with noise.
If the bandwidth of the detector is $10 \ \delta\nu_{\rm scint}$, then the maximum probability for such a scintillation event is $P_{\rm obs}\sim 7\times10^{-5}$. We present the probability as a function of S/N in the left panel of Fig. \ref{fig:P_scint}. One can see, as expected, that the probability is smaller for larger S/N. We also show the scintillation probability as a function of $n_1+n_2\equiv (\nu_2-\nu_1)/\delta\nu_{\rm scint}$ and $n_2 \equiv (\nu_{2b}-\nu_{1a})/\delta\nu_{\rm scint}$ in the right panel of Fig. \ref{fig:P_scint} for a fixed S/N of 5.

For the case (ii), when $\alpha$ is dependent on frequency, we consider the initial flux $f_0$ is described by a Gaussian function and $f_{\rm min}$ is still a constant value. The $\alpha_j$ in the $N$ channels ($j$ is from 1 to N) can be defined as
\begin{equation}
\alpha_j=\frac{f_{\rm min}}{A}{\rm exp}\left[\frac{(\nu_j-\nu_0)^2}{2\sigma^2}\right].
\end{equation}
In such case, we define $f_{\rm max,j}=S_N\alpha_j f_0$
Then the joint probability can be written as
\begin{equation}
P_{\rm obs} \simeq \prod_{j=1}^{n_1}P(0,f_{\rm min}) \prod_{j=1}^{n_2} P(f_{\rm min},f_{\rm max,j}).
\end{equation}

\begin{figure}
    \includegraphics[width=\columnwidth]{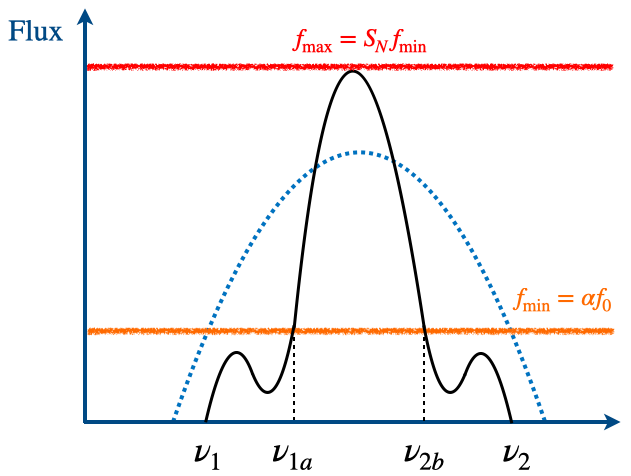}
    \caption{The blue dotted line is the original flux described by a Gaussian function. Orange line is the threshold flux of the detector. Black line is the observed flux due to the scintillation effect. Red line is the observed peak flux.}
    \label{fig:scint}
\end{figure}

\subsubsection{Observations of narrow bandwidth bursts}\label{sec:observation}
In this section, we discuss concrete examples of narrow FRB spectra and present the probability to make the spectrum narrow due to the scintillation effect.

\begin{itemize}
\item FRB 20190711 has the central frequency $\nu_0=1.4$ GHz and ${\rm FWHM}=65$ MHz, corresponding to $\Delta\nu/\nu_0\simeq0.05$ \citep{Kumar2021}. The bandwidth of the Parkes radio telescope is 0.7-4.0 GHz and no evidence of any emission in the remaining part of the 3.3 GHz band was observed. The integrated S/N can be roughly estimated as $\sim 5$. If this narrow spectrum is caused by scintillation, we have $\nu_{2}-\nu_{1}=3.3$ GHz and $\nu_{2b}-\nu_{1a}=65$ MHz can be regarded as the scintillation bandwidth $\delta \nu_{\rm scint}$. The probability due to scintillation can be estimated as $P_{\rm obs}\sim5\times10^{-8}$.
\item For FRB 20201124A, more than 600 bursts were detected by FAST with a bandwidth of 1.0-1.5 GHz. The histogram of bandwidth distribution has a peak at $\sim 0.28$ GHz. Let us consider a burst with such a narrow spectrum. Assume that the S/N has the typical value of $\sim 10$, the probability due that such a narrow spectrum is due to scintillation is $P_{\rm obs} \sim 0.06$.
\item For FRB 20220912A, a total of 1076 bursts were detected by FAST. Similarly, let us take an example case with $\Delta\nu/\nu_0 \sim0.2$ defined by the peak of its distribution, and the typical bandwidth would be at $181$ MHz \citep{ZhangYK2023}. Again assuming S/N $\sim 10$, the probability due that such a narrow spectrum is due to scintillation is $P_{\rm obs} \sim 0.007$.
\end{itemize}

\subsection{Filamentation instability of FRB pulse and its effect on the observed spectrum}\label{sec:filamentation}

An FRB pulse can break into filaments due to an instability that results from a combination of the ponderomotive force on charge particles and density dependence of the index of refraction. The fragmented pulse spreads laterally (diffractive spreading), and depending on their angular size, the observer could receive signals from multiple fragments. The interference of these signals would imprint a structure in the FRB spectrum. And, much like the scintillation bands, the amplitude of fluctuation with frequency is of order unity only when the lateral size of the FRB source is much smaller than the size of the fragments. For larger source sizes, the amplitude decreases as the ratio of fragment size to the source size increases. We provide an estimate of the fragment size using well-known results from linear analysis of the instability.

We make use of the results presented in \citet{Sobacchi2021} who have analyzed the filamentation instability as the EM wave moves through an electron-ion plasma. 
The maximum growth rate of the instability in electron-ion plasma is found to be
\begin{equation}
\Gamma_{\rm gr}^{ei}\simeq \frac{v_t^2}{v_e^2} \frac{\omega_{pe}^2}{\omega_{\rm frb}^2} \simeq\frac{a_0^2\omega_{pe}^2}{\omega_{\rm frb}},
 \label{growth-rate}
\end{equation}
where $\omega_{pe}=(4\pi e^2 n_{\rm e}/m_{\rm e})^{1/2}$ is electron plasma frequency,
\begin{equation}
a_0={eE_w\over m_{\rm e}c\omega_{\rm frb}}=\frac{eL_{\rm frb}^{1/2}}{m_ec^{3/2}\omega_{\rm frb} R}=0.5 \, L_{\rm frb,41}^{1/2} R_{13}^{-1} \nu_9^{-1}
\label{a0}
\end{equation}
is a dimensionless strength parameter for the FRB pulse, $E_w=\sqrt{L_{\rm frb}/(cR^2)}$ is the electric field strength associated with the isotropic FRB luminosity ($L_{\rm frb}$), $R$ is the distance from the central object where the plasma responsible for the instability resides, and $\omega_{\rm frb}$ is FRB wave angular frequency.

The transverse and longitudinal wave-numbers $(k_\perp, k_\parallel)$ of the fastest growing modes of this instability are
\begin{equation}
    c k_\perp \approx a_0 \omega_{pe},  \quad  c k_\parallel \approx \min\left\{ a_0\omega_{\rm frb}, \omega_{pe}\right\},
    \label{unstable-k}
\end{equation}

The angular size of fragments (asymptotically), after their lateral spreading, is
\begin{equation}
\theta_d \simeq \frac{c k_\perp}{\omega_{\rm frb}} \simeq \sqrt{ \frac{\Gamma_{\rm gr}^{ei}}{\omega_{\rm frb}}},
\end{equation}
where we made use of Eqs. (\ref{unstable-k}) and (\ref{growth-rate}). A faraway observer would receive photons from many different segments as long as $\theta_d > 2\pi/(k_\perp R)$, and thus a very thin pulse in the longitudinal direction would be temporally broadened to the duration
\begin{equation}
t_d \sim \max\left\{ t_{\rm FRB}, \, \frac{R \theta_d^2}{2c} \right\} \simeq \max\left\{ t_{\rm FRB}, \,\frac{R\Gamma_{\rm gr}^{ei}}{2c\omega_{\rm frb}} \right\}
\end{equation}
{due to the difference in arrival time of photons that have followed different geometrical paths as the pulse fragmented in the transverse direction and spread laterally due to wave diffraction; here $t_{\rm FRB}$ is the intrinsic duration of the FRB pulse.
The fragmentation of the FRB pulse in the longitudinal direction can cause short time variability when the width of the pulse is larger than $\sim 2\pi/k_\parallel$ or $t_{\rm FRB} > 2\pi/c k_\parallel$. }

{The interference of waves that arrive at the observer having traveled along different paths causes fluctuation of the observed spectrum. The frequency bandwidth for this fluctuation can be calculated in the same way as scintillation bandwidth, and is given by}
\begin{equation}
\Delta \nu_d\simeq {1\over 2\pi t_d} \simeq \frac{\omega_{\rm frb}^2 c}{\pi R a_0^2\omega_{pe}^2}.
\end{equation}
{The break-up of the FRB pulse in the longitudinal direction could also introduce fluctuations in the spectrum, but the resulting bandwidth would not be narrower than the intrinsic FRB spectrum, as per the Fourier theorem. }

\begin{figure*}
\begin{center}
\begin{tabular}{ll}
\resizebox{95mm}{!}{\includegraphics[]{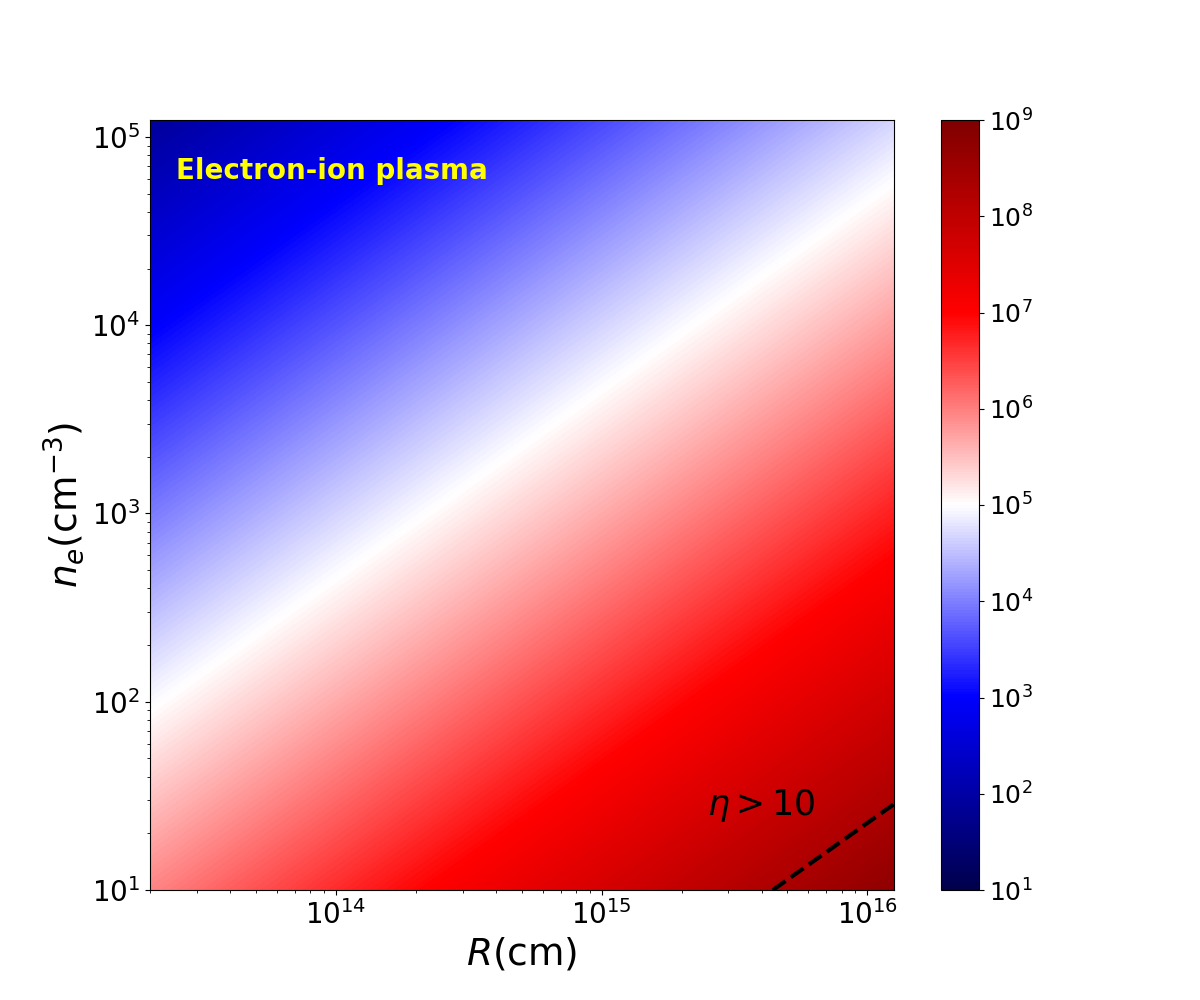}}&
\resizebox{95mm}{!}{\includegraphics[]{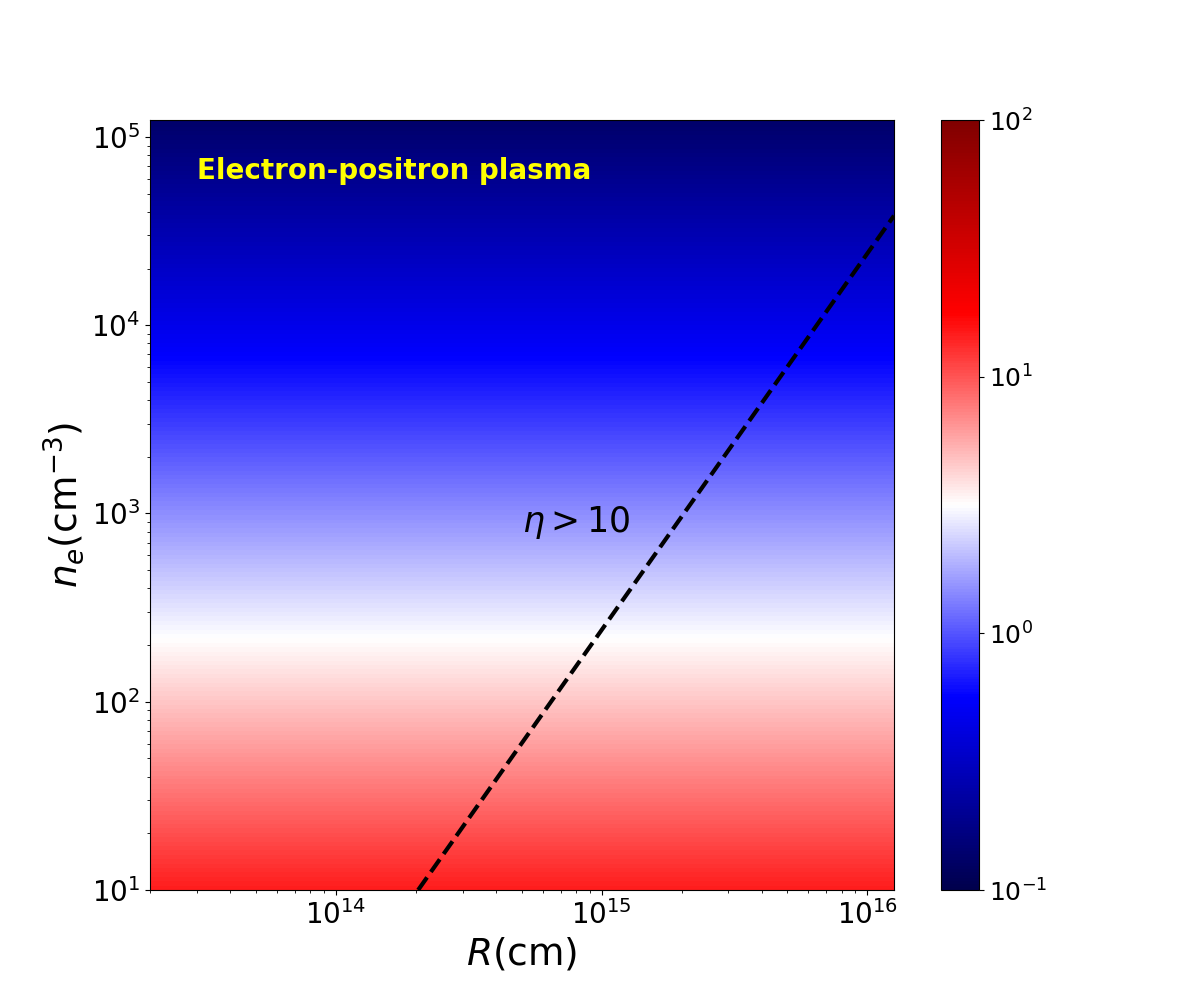}}
\end{tabular}
\caption{The frequency bandwidth due to the filamentation instability as a function of distance from the source ($R$) and the electron number density ($n_e$) for electron-ion plasma (left panel) and electron-positron plasma (right panel). The black dashed line denotes the e-folding growth of the filamentation instability $\eta=10$. The following parameters are adopted: FRBs luminosity $L_{\rm frb}=10^{42} \ {\rm erg \ s^{-1}}$,  frequency of FRBs $\nu_{\rm frb}=10^{9}$ Hz, and the duration $t_{\rm frb}=10^{-3}$ s. The color coding represents frequency bandwidth ($\Delta\nu_d$) in Hz. Note that $\Delta\nu_d$ is much smaller for e$^\pm$ plasma than for electron-ion plasma. Moreover, even for the electron-ion plasma, $\Delta\nu_d$ is much less than 100 MHz. Therefore, all fluctuations in the spectrum introduced by the filamentation instability would be wiped out when averaged over a few tens of MHz, and a broadband FRB pulse is highly unlikely to be converted into a narrowband pulse. Of course, if the lateral size of the source is much larger than the size of the fragments produced by the instability ($\ell_\perp$), then these fragments cannot cause any significant fluctuation in the spectrum, regardless of the value of $\Delta\nu_d$. The cutoff in $R$ below $2\times10^{13}$ cm is because the strength parameter $a_0>1$ below this radius and the calculation of the filamentation instability discussed in this work is not valid in that regime.  If the intrinsic FRB spectrum is broad, the probability that the filamentation instability is responsible for the observed narrow spectrum is very small, as quantified in  \S\ref{sec:scintillation} (see Eq. \ref{scint-prob4})  }
\label{fig:filamentation}
\end{center}
\end{figure*}

In the left panel of Fig. \ref{fig:filamentation}, we present the numerical results of the frequency bandwidth ($\Delta\nu_d$) as a function of distance to the central engine of FRBs ($R$) and the electron number density of the electron-ion plasma ($n_e$). 
The maximum frequency bandwidth can reach $\sim1$ MHz. 
{The distance of the plasma from the central engine is taken to be $R>2\times10^{13}$ cm to ensure that the EM wave strength parameter $a_0<1$ and the background plasma will not obtain the ultra-relativistic velocity, so that the results we have used for the filamentation instability in this sub-section are valid.}

From Eqs. (\ref{growth-rate}) \& (\ref{unstable-k}), it can be seen that the transverse size of fragmented FRB pulse is $\ell_\perp = 2\pi/k_\perp \sim (2\pi\lambda R)^{1/2}/\eta^{1/2} = \sqrt{2\pi}R_F/\eta^{1/2}$; where $\eta\equiv \Gamma_{\rm gr}^{ei}R/c$ is the total e-folding growth of the instability as the FRB pulse travels a distance $R$, and $R_{F}=\sqrt{\lambda R}$ is the Fresnel scale. Thus, a fully developed filamentation instability corresponds to the strong scintillation case, and they have the same effect on coherence bandwidth. When $\eta \gg 10$, the fragments are much smaller than the Fresnel scale, and as a result, the interference of waves from multiple fragments is observed. This means that the observed flux will vary by order of unity in amplitude over the bandwidth $\Delta\nu_d$, provided that the lateral size of the FRB source is much smaller than $R_F/\eta^{1/2} \sim R_F/5$. When the source size constraint is satisfied, the calculation of the probability that the instability might account for the observed narrow spectrum is the same as described in \S\ref{sec:scintillation}. On the other hand, when the source size is larger than $\sim R_F/5$, the flux variation amplitude would be much smaller over the bandwidth $\Delta\nu_d$; i.e., the filamentation instability would not be able to cause narrowing of the FRB spectrum.

The instability growth rate is larger if the FRB pulse travels through the $e^\pm$ wind from the magnetar. This is due to the additional contribution to the change in the index of refraction for radio waves associated with $e^\pm$ density perturbation. The transverse and longitudinal wavenumbers for the fastest growing modes of the filamentation instability when the FRB pulse travels in a cold, unmagnetized, $e^\pm$ medium are (e.g., \citealt{Sobacchi2023})
\begin{equation}
c k_\perp \sim  \left( a_0 \omega_{pe} \omega_{\rm frb}\right)^{1/2}, \quad \quad c k_\parallel \sim a_0 \omega_{pe},
\end{equation}
and the growth rate is
\begin{equation}
\Gamma_{\rm gr}^{\pm} \sim a_0 \omega_{pe}/\sqrt{2}.
\end{equation}
However, the instability can only grow for a time duration of order the FRB duration ($t_{\rm FRB}$), as opposed to the much longer time $R/c$, due to the fact that the density perturbation for e$^\pm$ only grows while the particles are moving through the pulse. 

The transverse size of the clumps that form when the instability fragments the FRB pulse is
\begin{equation}
    \ell_\perp \sim {2\pi\over k_\perp} \sim \left( {2\pi c t_{\rm FRB} \lambda_{\rm frb}\over \eta}\right)^{1/2} \sim 10^4 {\rm cm},
\end{equation}
where $\eta\equiv \Gamma_{\rm gr}^{\pm} \, t_{\rm FRB}$ is the total e-folding growth of the filamentation instability for e$^\pm$ plasma. If the transverse size of the FRB source is much larger than $\ell_\perp$, the amplitude of the observed flux variation observed with frequency due to the finite transverse size of the clumps formed by this instability would be much smaller than the mean flux. 

As before, the variability time imposed by the instability on the observed FRB lightcurve is
\begin{equation}
t_d \sim \max\left\{ t_{\rm FRB}, {R\over c} \left[ {\lambda k_\perp\over 2\pi}\right]^2
         \right\}.
\end{equation}
Or
\begin{equation}
t_d \sim \max\left\{  t_{\rm FRB}, \; {R\,\eta\over c (\omega_{\rm frb} t_{\rm frb})} \right\}.
\end{equation}
Thus, the observed spectrum develops structure, much like scintillation, with frequency bandwidth
\begin{equation}
\Delta \nu_d\simeq \frac{1}{2\pi t_d} \sim {25\, \rm Hz}\times \, {\nu_{\rm frb,9} \, t_{\rm FRB,-3}\over R_{15} \eta}.
\end{equation}
The filamentation instability reaches the nonlinear phase when $\eta \geq 10$, and so it is reasonable to take $\eta\sim 10$ for estimating $\Delta\nu_d$. Thus, we find the bandwidth to be no more than about 1 kHz. Therefore, if the intrinsic spectrum is broad, i.e. $\Delta\nu /\nu \sim 1$, there is a vanishingly small probability that the filamentation instability operating in an e$^\pm$ medium could be responsible for the observed narrow spectrum following the arguments presented in \S\ref{sec:scintillation}.
For instance, the probability is $\sim10^{-10^4}$ that an intrinsically broad-spectrum source is observed to have a bandwidth $\Delta\nu/\nu=0.1$, at 400 MHz with $S_N=5$ due to the filamentation instability (eq. \ref{scint-prob4}).

\section{Summary \& discussion}\label{sec:conclusion}

In this paper, we have investigated a variety of radiation mechanisms and propagation effects and studied the conditions required to produce the narrow spectra observed from some FRBs. The main conclusions of our study are summarized below.

There are two possible ways of making a narrow spectrum: one is through intrinsic radiation mechanisms; the other is through propagation effects (see Fig. \ref{fig:cladogram} for a summary). Relativistic Doppler effect places a severe constraint on the radiation mechanism capable of producing a narrow spectrum. 

We have shown that a generic constraint $\Delta\nu/\nu_0 > 0.58$\footnote{ \cite{Metzger2022} assumed that the bursts are intrinsically narrow and suggested that a single parameter, the power-law index of the frequency drift rate, can explain the narrow spectra and longer duration of repeating FRBs as well as the shorter time duration and wider spectra seen in non-repeating FRBs \citep{Pleunis2021}. 
In comparison, we consider monochromatic emission in the comoving frame of the FRB source and investigate the influence of high latitude emission on the observed spectra in this paper.} is placed on the intrinsic spectral width when the source is moving relativistically with Lorentz factor $\gamma$ and has an angular size $\theta_j$ that is larger than the Doppler beaming angle $\gamma^{-1}$. 

The angular size of the source for relativistic shocks far outside the magnetosphere has $\theta_j > \gamma^{-1}$ and is ruled out for FRBs with spectra where $\Delta\nu/\nu_0 < 0.58$. Magnetosphere models can have a source angular size much smaller then $\gamma^{-1}$ and are capable of producing very narrow spectra (See the discussion in \S\ref{sec:general}).

We have investigated several propagation effects, and found two in particular, viz. plasma lensing and scintillation, more promising than others. However, the probability of converting an intrinsically broad-spectrum source to a narrow spectrum with $\nu/\Delta\nu \gtrsim 5$ even for these mechanisms is less than 10$^{-3}$ when the signal-to-noise ratio is $\gtrsim 5$ (see eq. \ref{scint-prob4}).

High magnification plasma lensing events are highly chromatic. A broad spectrum radio pulse can be transformed into a narrow bandwidth wave after passing through a plasma lens, which highly magnifies the wave over a small frequency range $\Delta\nu$.
The electron column density required for a strong plasma lens is modest, on the order of about 10$^{12}$ cm$^{-2}$, but the density variation across the lens should be such that the difference in the phase shift of the wave along two trajectories through the lens almost exactly cancels out the phase shift due to the geometric path-length difference between these trajectories (see eq. \ref{density-needed}). Even a small fluctuation in the electron column density can disrupt this delicate balance, causing the screen to produce scintillation instead of plasma lensing. For instance, if the lens is a distance $R$ from the source, then a fluctuation in electron density of 10$^{-3}$cm$^{-3} R_{15}^{-1}$ across the Fresnel radius is sufficient to reduce or disrupt the lens magnification.
Moreover, large magnification is possible only for sources that have a size much smaller than the Fresnel radius corresponding to the distance between the source and the lens screen. This condition can be satisfied by a plasma lens that is close to the source. Thus, lensing may not explain repeaters with consistently narrow spectra over time as the lens would change with time (see \S \ref{sec:plasmalensing}). Moreover, if the narrow spectrum is due to lensing, the spectrum should have a predictable shape as given by Equation (\ref{plasma-mag4}).

Scintillation of an FRB is reflected in its spectrum as random flux variation with frequency, with a coherence bandwidth that is inversely proportional to the scattering time in the turbulent plasma screen responsible for the scintillation. Considering the random nature of scintillation-caused flux variation, there is a non-zero probability that the waves scattered by different parts of the screen interfere destructively outside a narrow frequency range of width $\Delta\nu$, thereby manifesting to an observer as a source with a small bandwidth. The probability for this is presented in \S\ref{sec:scintillation}. For an FRB spectrum with $\nu/\Delta\nu \gtrsim 5$, this probability is too small for scintillation to be a viable mechanism for the narrowness of the spectrum. Moreover, scintillation cannot explain an FRB where the transverse size of the radio source is larger than the diffractive length of the scattering screen. Since scintillation bandwidth increases with wave frequency as $\sim\nu^4$, we should expect $\Delta\nu$ to increase similarly if the narrowness of the observed spectrum is due to scintillation. This prediction should be verified whenever data covering a large dynamical range in frequency is available modulo the fact that the intrinsic spectrum might have a width $\Delta\nu\sim\nu$.

The well-known filamentation instability, operating within 0.1 pc of the source, can break up a high-amplitude FRB radio pulse into filaments, imposing fluctuations in the observed flux with frequency much like scintillation. This mechanism shares the weaknesses of the scintillation mechanism we have described, as well as its own limitations related to the growth rate of the instability and the size of filaments produced by it (see \S\ref{sec:filamentation}). {In particular, we emphasize that if the intrinsic FRB spectrum is broad, the probability that the filamentation instability is responsible for the observed narrow spectrum is very small, as quantified by eq. \ref{scint-prob4} in \S\ref{sec:scintillation}. }

The far-away class of FRB models, where coherent radio emission is produced outside of the light-cylinder (LC), share a common feature: the relativistic outflow originating from the central compact object, which is ultimately responsible for the energy in the burst, is expected to have an angular size much larger than $\gamma^{-1}$ due to diverging magnetic field lines that extend outside the LC and the lateral expansion of the jet. Therefore, this class of FRB models would produce a spectrum with a bandwidth $\Delta\nu/\nu \gtrsim 0.5$. In particular, the synchrotron maser mechanism operating near the shock front has roughly the same growth rate over a bandwidth $\Delta\nu\sim \nu$. Therefore, even without the relativistic Doppler effect, we expect the emergent spectrum to not be narrow. Moreover, the lateral size of the visible part of the source for this class of FRB models is $R/\gamma \gtrsim 10^{9}$ cm, which exceeds the Fresnel scale for plasma screens within 0.1 pc of the source; $R$ is the distance from the central object where the radio emission is produced. Consequently, propagation effects such as plasma lensing, filamentation instability, or scintillation cannot transform a broadband spectrum produced in far away class of models into a narrowband signal.

In the appendix we have discussed the possibility of coherent curvature radiation by bunches (\S\ref{Appendix:radiation}), coherent inverse Compton scattering of low frequency waves by particle bunches (\ref{sec:intrinsic}), resonant Cherenkov radiation (\S\ref{Appendix:Cherenkov}), and absorption of radio waves by resonant cyclotron
absorption inside magnetosphere and synchrotron absorption (\S\ref{Appendix:absorption}), and find that they require either very narrow distribution of particle momentum vector or some other finely tuned condition in order to convert a broadband intrinsic source spectrum into a narrow band pulse.

The main conclusion of this work is that narrow FRB spectra with $\nu/\Delta\nu \gtrsim 2$ are likely an intrinsic property of the source if the signal to noise for the spectrum is large ($\gtrsim 5$). In some cases, a narrow-band spectrum could result from scintillation or plasma lensing when the signal-to-noise ratio of the data is less than about 5. We have discussed the probability of propagation effects converting a broad-band signal to a narrow spectral event in \S\ref{sec:scintillation}.

\section*{Acknowledgements}
PK’s work was funded in part by an NSF grant AST-2009619 and a NASA grant 80NSSC24K0770. YQ acknowledges a Top Tier Doctoral Graduate Research Assistantship (TTDGRA) at University of Nevada, Las Vegas. BZ acknowledges NASA grant 80NSSC23M0104 and Nevada Center for Astrophysics for support. PK is grateful to Ramesh Narayan and Ue-Li Pen for very helpful discussion about scintillation and plasma lensing.

\appendix

\section{RADIATION MECHANISMS FOR FRBS}

\begin{itemize}
\item Curvature radiation: Charged bunches moving along curved strong magnetic field lines radiate coherent curvature radiation. This mechanism to power FRBs has been discussed by many authors \citep[e.g.][]{katz2014,Kumar2017,Yang&zhang2018,Kumar&Bosnjak2020,Lu20,Cooper2021,Kumar22,Wang2022,Wang2022b,QZK}, with the requirement that a parallel electric field ($E_\parallel$) exists to balance the radiation cooling and acceleration of the emitting bunches \citep{Kumar2017,Kumar&Bosnjak2020}. The curvature radiation of a single particle has a broad spectrum \citep{Rybicki&Lightman1979,Jackson1998} (Appendix). In order to achieve narrow spectra, specific arrangements of emitting bunches in space and time are required, either via charge separation \citep{Yang2022} or allowing equal spacing between charged bunches \citep{Yang2023,wang2023}\footnote{Narrow spectra caused by a special spatial arrangement of bunches are defined as ``coherent narrow'', in contrast to ``intrinsic narrow'' which is related to the radiation mechanism itself.}. However, it is unclear how physically the bunches can be arranged with the desired patterns. For this reason, we mark curvature radiation in orange color under ``coherent narrow'' in Figure \ref{fig:cladogram} and  will not discuss this mechanism further. 
\item Inverse Compton scattering: FRBs may be associated significant quakes at the magnetar crust region. Such osccilations can generate fast magnetosonic (X-mode) waves which propagate freely in the magnetosphere. 
Such low-frequency electromagnetic waves may be upscattered by relativistic, charged bunches and produce FRBs through the inverse Compton scattering (ICS) mechanism \citep{Zhang22,QZK,Qu&Zhang2024}. If the frequency of the incident low-frequency waves is narrow (which may correspond to a certain oscillation mode in the crust), the mechanism can produce narrow spectra because the radiation mechanism itself does not have a broad spectrum and that the bunches could be coherently enhaced. We discuss this mechanism in detail in Sect \ref{sec:intrinsic}.
\item Cyclotron/Synchrotron radiation: These mechanisms have been introduced to interpret FRBs within the synchrotron maser models that invoke electron gyration in ordered magnetic fields in quasi-perpendicular shocks 
\citep{Lyubarsky2014,Metzger2019, Beloborodov2020,Plotnikov&Sironi2019}. Numerical simulations show that the narrowness of the spectrum can reach $\Delta\nu/\nu_0\sim0.2$ \citep{Sironi2021} in the case that electrons are non-relativistic. This is because cyclotron radiation is intrinsically line emission. When electrons are relativistic, the spectrum becomes broader because the introduction of the modified Bessel function for synchrotron radiation. We discuss these mechanisms in the Appendix. The synchrotron maser model is subject to the generic geometric constraint as discussed in the Sec. \ref{sec:general}, so the spectrum cannot be narrower than $\Delta\nu/\nu_0\simeq0.58$.
\item Cherenkov radiation: When charged particles propagate faster than the phase light speed in the background plasma with a refractive index $n_r>1$, radiation can be generated even if the particles have no acceleration. 
When the velocity of the plasma equals the wave phase velocity, Cherenkov instability would occur with the growth rate significant at frequencies higher than GHz \citep{Lu&Kumar2018}. 
Recently, \cite{Liu2023} proposed that FRBs might be produced through Cherenkov radiation of ``superluminal'' particles
the collective plasma emission is in a specific frequency range. We investigate this mechanism in the context of FRBs and 
discuss some drawbacks of the mechanism in Appendix \ref{Appendix:Cherenkov}.
\end{itemize}

\subsection{Cyclotron/Synchrotron/Curvature radiation}\label{Appendix:radiation}

Non-relativistic electrons gyrating in magnetic fields produce cyclotron radiation with a typical angular frequency $\omega'_B= e B'/mc$ in the comoving frame of the emitter. The spectrum of cyclotron maser emission is intrinsically narrow since the emission power of higher harmonics drops quickly and cannot contribute significantly to the emission power. In the observer frame, the frequency is boosted by the Doppler factor as
$\omega_B={\cal D}\omega_B'$. Because of the high-latitude effect (as discussed in Sect. \ref{sec:general}), the observed spectrum is wider than $\Delta \nu / \nu \sim 0.58$.

For relativistic particles moving in a magnetic field, 
the emission spectrum can be written as \citep{Rybicki&Lightman1979,Jackson1998}
\begin{equation}
\frac{d^2W}{d\omega d\Omega}=\frac{e^2}{3\pi^2c}\left(\frac{\omega\rho}{c}\right)^2\left(\frac{1}{\gamma^2}+\theta^2\right)^2\left[K_{2/3}^2(\xi)+\frac{\gamma^2\theta^2}{1+\gamma^2\theta^2}K_{1/3}^2(\xi)\right],
\end{equation}
where $\rho$ is the gyration radius of particles for synchrotron radiation and the curvature radius of the field lines for curvature radiation. One notices that the spectrum is intrinsically broad due to the existence of the modified Bessel function, which originates from the fact that the acceleration direction is perpendicular to velocity.
In order to explain the observed narrow spectrum, one must consider 
many bunches that are spatially coherent so that narrow spectra are obtained through coherent phase superposition \citep{Yang2023,wang2023}.

\subsection{Coherent inverse Compton scattering by charged bunches}\label{sec:intrinsic}

For models invoking magnetospheric emission, the radiation mechanism itself needs to produce a narrow spectrum in order to make an observed narrow spectrum. Coherent curvature radiation makes an  intrinsically broad spectrum with $\Delta\nu/\nu>1$ (Appendix \ref{Appendix:radiation}). Here we consider coherent inverse Compton scattering (ICS) mechanism, which has the merit of making intrinsically narrow spectra \citep{Zhang22,Qu&Zhang2024}. Within this picture, FRBs could be produced by charged bunches that scatter off the low-frequency X-mode electromagnetic waves (or fast magnetosonic waves, i.e. F-mode) with $\nu=10$ kHz, which may be generated through near-surface plasma oscillations induced by magnetar crust cracking. 
The incident waves act as a perturbation which has a perpendicular electric field component with respect to the background magnetic field to accelerate the charged bunches. We consider that electrons can only move along the background strong magnetic field lines and their  motion equation in the co-moving frame can be written as
\begin{equation}
m_e\frac{d^2\vec{r}'}{dt'^2}=e\Vec{E}'+\frac{e}{c}\left(\frac{d\vec{r}'}{dt'}\times\Vec{B}'\right).
\end{equation}
The ICS spectrum of one single electron in the comoving frame of the electron can be written as \citep{Rybicki&Lightman1979,Jackson1998}
\begin{equation}
\begin{aligned}
\frac{d^2W'}{d\omega'd\Omega'}=\frac{e^2\omega'^2}{4\pi^2c}\left|\int_{-\infty}^{+\infty}{\hat{n}'\times(\hat{n}'\times{\vec \beta'})}{\rm exp}([i\omega'(t'-\hat{n}'\cdot\vec r'/c)]dt' \right|^2,
\end{aligned}
\end{equation}
where $\hat{n}'$ and $\vec \beta'$ are the line-of-sight vector
and dimensionless velocity, respectively, in the comoving frame.

In order to obtain the radiation spectrum of a relativistically-moving electron, we perform the Doppler transformation of the spectrum as
\begin{equation}
\frac{d^2W}{d\omega d\Omega}={\cal D}^{2}\frac{d^2W'}{d\omega' d\Omega'}.
\end{equation}
The fundamental reason is that the incident low-frequency waves are monochromatic, so that the scattered waves are also nearly monochromatic. 
If the incident low frequency waves are not monochromatic with a frequency bandwidth $\Delta\nu_i$, the scattered waves bandwidth would be enhanced by a factor of $\sim \gamma^2\Delta\nu$, one can see that the ratio $\Delta\nu/\nu_0$ is invariant for both incident waves and scattered waves.

\subsection{Cherenkov radiation}\label{Appendix:Cherenkov}
Cherenkov radiation can happen in a medium with refractive index $n_r>1$. In such a case, the speed of charged particles can exceed the phase velocity of light. The Lienard–Wiechert potential of an accelerating charged particle should be modified by replacing $c \rightarrow c/n_r$. Thus Cherenkov radiation bandwidth is determined by $n_r^2>1/\beta^2$. The denominator in Lienard–Wiechert potential is modified by $\kappa=1-n_r\beta\cos\theta$. One can see that $\kappa$ can be zero as the resonance condition for a specific angle $\cos\theta_c=1/({n_r\beta})$ denoting the main radiation direction which is related to the refractive index depending on wave frequency and the radiation energy is mainly at the resonant frequency.

For the case of electrostatic waves, the electric field is parallel to the propagation direction (the background magnetic field $\vec B_0$ is along $z$-axis) and we have $\vec k\times\vec B_0=0$. In such a case, there is a resonance when the particle velocity equals the parallel phase velocity of the wave, i.e.
\begin{equation}
\omega-k_\parallel v_\parallel=0.
\end{equation}
where the subscript $\parallel$ denotes the components parallel to background magnetic field.
For the magnetized plasma case, the resonance condition is
\begin{equation}
\omega-k_\parallel v_\parallel-s\omega_B=0,
\end{equation}
where $s=0,\pm 1,\pm 2,...$ is the harmonic number. 
The spectrum bandwidths $\Delta\nu$ is determined by particles’ velocity and magnetic fields.
In order to calculate the Cherenkov radiation spectrum, one can replace the speed of light $c\rightarrow c/n_r$ since the Lienard–Wiechert potential of a charged particle is modified in the dispersive medium. 
It is unlikely that the plasma maser mechanism is responsible for FRBs:
(i) The fundamental difficulty to allow the plasma maser mechanism is that the growth rate for beam instabilities is negligible in the radio band \citep{Lu&Kumar2018}. 
(ii) In order to produce coherent radiation via Cherenkov mechanism, one has to consider collective plasma distributed in a relatively narrow spatial scale compared with typical FRB wavelength, i.e. in charged bunches. The acceleration of the bunch due to a large scale parallel electric field is required. In such a case, the background plasma surrounded by the bunch would be charge-separated and the efficiency for such a mechanism (which requires the presence of the plasma) would be reduced.

\section{Absorption processes}\label{Appendix:absorption}

In this Appendix, we discuss two possible absorption effects that may influence the FRB spectrum: resonant cyclotron absorption and synchrotron absorption by relativistic electrons. We consider an initial incident spectrum with intensity $I_0$. Then the escaped spectrum after a absorption process can be written as
\begin{equation}
I=I_0{\rm exp}(-\tau),
\end{equation}
where $\tau\simeq \alpha L$ is the optical depth of absorption process, $\alpha$ is the absorption coefficient of corresponding physical processes and $L$ is the typical length scale. In the following, we assume that the original spectrum is described by a Gaussian function.

\subsection{Resonant cyclotron absorption}

The magnetar magnetosphere is filled with an electron-positron plasma, likely moving relativistically along background strong magnetic field in the open field line region. In the rest frame of electrons, FRB waves will be absorbed via resonant cyclotron absorption when the incident waves frequency is exactly equal to the cyclotron frequency, i.e
\begin{equation}
\omega'=\gamma_{\pm}\omega_{\rm frb}(1-\beta\cos\theta_B)=\omega_B,
\end{equation}
where $\omega'$ is the FRB wave angular frequency in the rest frame of lepton, $\gamma_+$ and $\gamma_-$ are the Lorentz factors of positrons and electrons, respectively, and $\theta_B$ is the angle between the wave vector and the magnetic field at the resonance radius. This effect will cause the spectrum to have a sharp absorption line, but it cannot convert a broad FRB spectrum to be narrow.

\subsection{Synchrotron absorption}
\begin{figure}
	\includegraphics[width=\columnwidth]{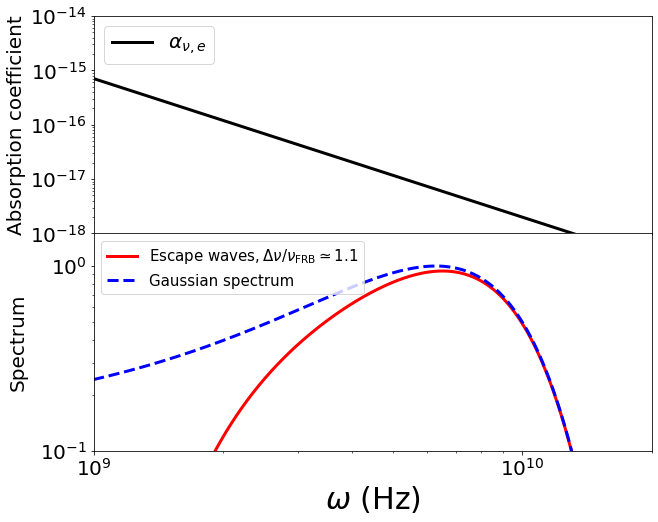}
    \caption{Gaussian spectrum undergoes synchrotron absorption in supernova remnant. Following parameters are adopted: nebula radius and absorption length scale $r=10^{17}$ cm, absorption region $\Delta r=10^{16}$ cm, specific synchrotron luminosity of the nebular $L_{\rm \nu, nebula}=10^{29} \ \rm erg \ s^{-1} \ Hz^{-1}$, FRB typical frequency $\nu_{\rm frb}=10^9$ Hz, magnetic field $B=10^{-3}$ G and the power-law index $p=1.1$.}
    \label{fig:syn absorption}
\end{figure}

Three repeating FRBs, i.e. 20121102A \citep{Chatterjee2017,Marcote2017}, FRB 20190520B \citep{Niu2022}, and FRB 20201124A \citep{bruni2023} are found to be each associated with a persistent radio source (PRS). It is suspected that the these FRB sources have an associated synchrotron-emitting PRS, possibly a supernova remnant, a magnetar wind nebula, or a mini-AGN. When the incident FRB photon frequency is exactly equal to the synchrotron radiation frequency by the electrons in the PRS region, i.e. $\nu_{\rm frb}=\nu_{\rm syn}$, the spectrum of the escaped waves would have a low frequency cut off below the synchrotron absorption frequency. The absorption coefficient for synchrotron radiation is given by \citep{Rybicki&Lightman1979}
\begin{equation}
\alpha_\nu=-\frac{1}{8\pi\nu^2m_e}\int_{\gamma_{\rm min}}^{\gamma_{\rm max}}d\gamma P(\gamma,\nu)\gamma^2\frac{\partial}{\partial \gamma}\left[\frac{N(\gamma)}{\gamma^2}\right],
\end{equation}
where $P(\gamma,\nu)$ is the specific synchrotron radiation power. 
We consider a relativistic electron gas with a power-law distribution in Lorentz factor, i.e.
$N(\gamma_e)d\gamma_e=C_{\gamma_{e}}\gamma_e^{-p}d\gamma_e$ with $\gamma_{\rm min}<\gamma_e<\gamma_{\rm max}$ and $p>1$. The total electron number density can be calculated as
\begin{equation}
n_{e}=\int_{\gamma_{\rm min}}^{\gamma_{\rm max}}C_{\gamma_e}\gamma_e^{-p}d\gamma_e=\frac{C_{\gamma_{e}}}{p-1}(\gamma_{\rm min}^{-p+1}-\gamma_{\rm max}^{-p+1}).
\end{equation}
We consider a specific case that the radius of the magnetar wind nebula is $r=10^{18}$ cm, the region length scale is $\Delta r=10^{17}$ cm and the magnetic field strength is chosen as $B=10^{-3}$ G.
We apply the specific form of total power $P(\gamma,\nu)$ for synchrotron radiation, the absorption coefficient can be written as
\begin{equation}
\alpha_{\nu,e}=\frac{p+2}{8\pi m_e}C_{\gamma_{e}}\nu^{-2}\int_{\gamma_{\rm min}}^{\gamma_{\rm max}}\frac{\sqrt{3}e^2B_\perp}{m_{e}c^2}F(x)\gamma_{e}^{-(p+1)}d\gamma_{e},
\end{equation}
where 
\begin{equation}
    F(x)=x\int_{x}^{\infty}K_{5/3}(\xi)d\xi\sim\left\{
    \begin{aligned}
    &\frac{4\pi}{\sqrt{3}\Gamma(1/3)}\left(\frac{x}{2}\right)^{1/3}, &x\ll 1,\\
    &\left(\frac{\pi}{2}\right)^{1/2}x^{1/2}e^{-x}, &x\gg 1
    \end{aligned}
\right.
\end{equation}
describes the synchrotron spectrum of a single particle in a uniform magnetic field, $\Gamma(1/3)$ is the gamma function of argument $1/3$, $x=\omega/\omega_{\rm ch,frb}=\nu/\nu_{\rm ch,frb}$ and $\nu_{\rm ch,frb}=\omega_{\rm ch,frb}/(2\pi)=3\gamma_{\rm ch,frb}^2eB_\perp/(4\pi m_ec)$ is the characteristic synchrotron emission frequency. 
We consider the case when $\gamma_{\rm min}\ll\gamma({\nu_{\rm ch,frb}})\ll\gamma_{\rm max}$. The absorption coefficient can be integrated as
\begin{equation}
\begin{aligned}
&\alpha_{\nu,e}=\frac{\sqrt{3}e^3C_{\gamma_e}}{8\pi m_e^2c^2}\left(\frac{3e}{2\pi m_ec}\right)^{p/2} B_\perp^{\frac{p+2}{2}}\Gamma\left(\frac{3p+2}{12}\right)\Gamma\left(\frac{3p+22}{12}\right)\nu^{-\frac{p+4}{2}}\\
&\simeq[10^4(8.4\times10^6)^{\frac{p}{2}} \ {\rm cm^{-1}}]C_{\gamma_e} B_\perp^{\frac{p+2}{2}}\Gamma\left(\frac{3p+2}{12}\right)\Gamma\left(\frac{3p+22}{12}\right)\nu^{-\frac{p+4}{2}}.
\end{aligned}
\end{equation}
We present the absorption coefficient as a function of angular frequency in the upper panel of Fig. \ref{fig:syn absorption}. One can see that the absorption effect is dominant in the low frequency tail. We also present the original Gaussian-form spectrum (blue solid line) and the escaping wave spectrum (red dashed line) after synchrotron absorption as a function of angular frequency in the lower panel of Fig. \ref{fig:syn absorption}. We conclude that such an effect is not responsible for producing narrow spectra in FRBs because the observed low cut-off frequencies are different from burst to burst.

\end{document}